\documentclass{emulateapj}

\newcommand{\ceio}{C$^{18}$O}
\newcommand{\thco}{$^{13}$CO}

\newcommand{\hco}{HCO$^+$}
\newcommand{\co}{$^{12}$CO}
\newcommand{\kms}{km s$^{-1}$}

\shorttitle{3 mm Spectra of Galaxies}
\shortauthors{Snell et al.}

\begin{document}

\title{The Redshift Search Receiver 3 mm Wavelength Spectra of 10 Galaxies}

\author{Ronald L. Snell\altaffilmark{1}, Gopal Narayanan\altaffilmark{1},
Min S. Yun\altaffilmark{1}, Mark Heyer\altaffilmark{1},
Aeree Chung\altaffilmark{1,2}, William M. Irvine\altaffilmark{1,3}, Neal R. Erickson\altaffilmark{1},
and Guilin Liu\altaffilmark{1}}
\altaffiltext{1}{Department of Astronomy, LGRT 619, University of Massachusetts,
710 North Pleasant Street, Amherst, MA 01003; emails: snell@astro.umass.edu, gopal@astro.umass.edu
myun@astro.umass.edu, heyer@astro.umass.edu, irvine@astro.umass.edu, neal@astro.umass.edu,
gliu@astro.umass.edu}
\altaffiltext{2}{Harvard-Smithsonian Astrophysical Observatory, 60 Garden Street, Cambridge,
MA, 02138; email: achung@cfa.harvard.edu}
\altaffiltext{3}{Goddard Center for Astrobiology}

\begin{abstract}

The 3 mm wavelength spectra of 10 galaxies have been
obtained at the Five College Radio Astronomy Observatory using
a new, very broadband receiver and spectrometer, called the Redshift 
Search Receiver (RSR).  The RSR has an instantaneous bandwidth of
37 GHz covering frequencies from 74 to 111 GHz, and has a spectral 
resolution of 31 MHz ($\sim$100 km s$^{-1}$).  
During tests of the RSR on the FCRAO 14 m telescope the complete 
3 mm spectra of the central regions of
NGC 253, Maffei 2, NGC1068, IC 342, M82, NGC 3079, NGC 3690, NGC 4258, Arp 220 and
NGC 6240 were obtained.  Within the wavelength band covered by the RSR, 
20 spectral lines from 14 different atomic and molecular species were detected.  
Based on simultaneous fits to the spectrum of each galaxy, a number
of key molecular line ratios are derived.  A simple model which assumes the emission
arises from an ensemble of Milky Way-like Giant Nolecular Cloud cores can adequately fit
the observed line ratios using molecular abundances based on Galactic molecular cloud cores.
Variations seen in some line ratios, such as \thco/HCN and HCO$^+$/HCN, can be
explained if the mean density of the molecular gas varies from galaxy
to galaxy.  However, NGC 3690, NGC 4258 and NGC 6240 show very large HCO$^+$/HCN
ratios and require significant abundance enhancement of HCO$^+$ over HCN, possible
due to the proximity to active galactic nucleus activity.  Finally, the mass of dense 
molecular gas is estimated and we infer that 25-85\% of the total molecular gas in
the central regions of these galaxies must 
have densities greater than 10$^4$ cm$^{-3}$.

\end{abstract}


\section{Introduction}

The molecular emission lines observed in galaxies are powerful probes of the
physical and chemical properties of the gas most directly connected to the
star formation process.  Over the past several decades there have been many 
studies of the molecular line emission from galaxies, and these studies have
provided invaluable information on the molecular gas content, star formation 
efficiency and molecular abundances \citep{you91, omo07}.  
Although CO emission has been the primary means of deriving the molecular gas content 
in galaxies, emission from other less abundant molecular species, 
is much better in determining the physical and chemical properties of the gas.  
As the chemistry of the molecular gas is affected by the local radiation field, 
these trace molecular constituents may also provide information on local environments 
of this gas.  More frequently observations of molecular species such as HCN, 
HCO$^+$, HNC, and CS are being obtained \citep{omo07} and in a few 
cases mapped in nearby galaxies \citep{meie05}.  Since 
these molecules have much larger permanent electric dipole moments than CO, they 
require considerable higher densities to collisionally excite, and thus, in 
general, trace denser molecular gas than the gas probed by CO.  Emissions
from these dense gas tracers, such as HCN, have been argued to be a much better 
probe of the star-forming molecular gas than emission from CO \citep{gao04}.

The utility of emission from these `high dipole moment molecules' to probe
the properties of the nuclear regions of galaxies is well established, and there
have been numerous papers on this subject in recent years (see
review by \citet{omo07}).  In addition, several
emission line ratios, such as those for the low-lying transitions of
HCN/CO, HCO$^+$/HCN, and HNC/HCN, have been suggested as
good diagnostics of the properties of the dense star-forming gas in galaxies
\citep{aal08}.  Nearly all studies of the molecular emission in galaxies have 
been limited to just a few targeted molecular lines.  However, \citet{mar06} recently 
presented the first spectral scan of a galaxy providing an inventory of the molecular
lines in NGC 253 within the 2 mm wavelength band.  Such spectral scans can 
provide a much more complete description of the chemical complexities of the molecular 
gas in galaxies.  The 3 mm wavelength band is equally well suited,
and past spectral surveys have 
provided important information on the chemical and physical properties of 
Giant Molecular Clouds (GMCs) in the Milky Way \citep{joh84, joh85, cum86, tur89}.
One major disadvantage of most spectral line surveys is that the data are assembled
from many observations with varying pointing accuracy and with potentially systematic
calibration problems.  The observations described here have been obtained 
{\it simultaneously} over the full spectral band for each galaxy, and hence
many systematic problems are eliminated.  
In this paper we present the first 3 mm spectral scans of the central regions
of 10 galaxies.

\section{Observations and Data Reduction}

The observations reported in this paper were obtained in spring 2008
using the Redshift Search Receiver (RSR) on the Five College Radio Astronomy
Observatory (FCRAO) 14-m telescope.  The RSR is a sensitive, ultra-broad bandwidth
receiver/spectrometer \citep{eri07} developed at the University of
Massachusetts as a facility instrument for the 50-m diameter
Large Millimeter Telescope \citep{sch08}.  This instrument was designed primarily to 
measure the redshift of distant, dust-obscured galaxies.  The RSR is a 
dual polarization and dual beam instrument.  The four broadband receivers
cover instantaneously the frequency range 74-111 GHz.  A high 
speed Faraday rotation beam switch, operating at 1 kilohertz, is used 
to overcome the 1/{\it f} noise originating in the front-end monolithic microwave
integrated circuit (MMIC) amplifiers.  Following the MMIC amplifiers, two
wideband mixers convert each receiver band to two intermediate
frequency (IF) channels.  After further conversion and amplification,
the IF signal is passed into an analog auto-correlation spectrometer.  Each
analog correlator has a bandwidth of 6.5 GHz and there are six correlators
for each receiver polarization. To obtain the best frequency resolution, we
do not apodize the lag domain signal before transforming to the frequency
domain.  Without this apodization, as with other auto-correlation based 
spectrometers, a ringing effect can be seen in the baselines around strong,
narrow lines.  The RSR has been designed to detect weak relatively broad
lines, in which case this ringing is not a problem.
In several of the spectra presented here, the \thco\ line
is sufficiently strong, that this ringing adds additional
baseline noise near the line.  The RSR has an instantaneous 
bandwidth of 37 GHz with a resolution
of 31 MHz, or a velocity resolution of approximately 100 \kms\ in 
the 3 mm wavelength band.  

The RSR was commissioned on the FCRAO 14-m telescope in 2007 and 2008
and used for several initial science projects.  During the time of
the observations reported here, 12 of the final 24 spectrometers were
available, which permitted beam switching with the full 37 GHz bandwidth
in a single polarization.  During commissioning observations, there was a small
hardware issue (which has since been diagnosed and fixed) which produced anomalous
noise at approximately 92 GHz, and a small region around this frequency has been 
blanked in many of the spectra presented here.

Observations were obtained at the positions of the 10 galaxies listed 
in Table 1.  Distances in Table 1 for the nearby galaxies,
NGC 253, Maffei 2, IC342 and M82, are from \citet{kar05} and distances
for the more distant galaxies are from the NASA/IPAC 
Extragalactic Database.  The galaxies were selected primarily because 
they were well studied and had relatively bright molecular emission lines,
important for verifying the performance of the RSR.  Since our observations 
cover the entire 3 mm wavelength window with uniform sensitivity, we can study 
the previously detected lines and all other lines within frequency range
of the RSR.  We added to the list of bright line galaxies some additional
weak-line galaxies to further test the performance of
the RSR and many of these additional galaxies are known to host an 
active galactic nucleus (AGN).  Also included in the table is the total 
integration time spent on each galaxy.  The observations were taken over
varying atmospheric conditions, so our broad aims were to achieve
a relatively uniform sensitivity of about 1 mK or to achieve a signal to
noise in the $^{13}$CO line of 10.  NGC 3079 and NGC 6240 did not
show strong lines, so we spent additional time on these sources
to lower the noise to better than 0.5 mK.  The observations for each
galaxy were obtained over several observing sessions and combined
together.  The pointing and calibration was repeatedly checked by
observations of the continuum emission from planets and quasars.  Pointing
offsets were never more than a few arcseconds, a small fraction of the beam
size, and the overall flux calibration is repeatable to be better than 10 \%. 

Also included in Table 1 are some brief notes regarding the properties of 
the central regions of the galaxies in our sample. The molecular 
emission in many of these galaxies
is dominated by their nuclear starburst.  Both NGC 6240 and 
Arp 220 are ultra luminous infrared galaxies (ULIRGS).  NGC 1068, 
NGC 4258 and NGC 6240 all have Seyfert 2 nuclei; however, the emission
we observe in NGC 1068 is likely dominated by the surrounding nuclear 
starburst ring \citep{sch00}.  NGC 4258 is a weak AGN, but has a pair of
radio jets that may be influencing the molecular emission \citep{kra07}.
NGC 6240 is the merger of two Seyfert 2 host galaxies where the two
AGNs are separated by less than 2\arcsec\ and the molecular emission
is concentrated in a small region centered on the Seyfert nuclei \citep{ion07}.
Finally we note that NGC 3690 is in the process of merging
with IC 694 (the system is called Arp 299 or Mrk 171).  The nuclei of
NGC 3690 and IC 694 are separated by only $\backsim$ 20 arcseconds and
both nuclei have an AGN \citep{gar06}. The
\thco\ and HCN emission from the IC 694 nucleus is much stronger than 
that from the NGC 3690 nucleus \citep{aal97}; thus, although we are centered on 
NGC 3690, the emission may have a contribution from IC 694.  Unfortunately,
the emission from both nuclei is at approximately the same velocity, so
velocity cannot be used to separate the emissions within our telescope
beam.

One limitation of the RSR for the observations presented in this
paper is the relatively small 
throw of the beam switch.  The reference beam is only offset by 4.34
arcmin in azimuth.  Thus, for the largest 
galaxies, such as IC 342 or NGC 253, the reference beam may contain weak 
emission from molecular clouds in the galactic disk.  However, the \co\ 
emission falls off sharply from the central regions, and even in the largest 
galaxies, the emission in the reference beam is more than 10 times 
weaker than that at the center of the galaxy \citep{you95}.  
As a check on line fluxes we compared the integrated intensity of the 
\thco\ emission, which of the lines we observe has the greatest potential
for disk contamination, in the four galaxies (IC 342, NGC 253, NGC 1068 and NGC 3079)
that were in common with the study of \citet{pag01}. \citet{pag01} also
used the FCRAO 14 m telescope; however they obtained their observations by 
position switching well off the galaxy.  We find in all five galaxies that
the integrated intensity we measure is in good agreement
with that of \citet{pag01}; thus, we believe that the line integrated 
intensities that we report have not been strongly affected by our beam 
switching observation mode.  

The data have been analyzed using software developed by G. Narayanan 
specifically for RSR data reduction, and this software is a complete
data reduction package with interactive graphics implemented using the
Python programming language.  For the data for each galaxy, linear 
baselines were subtracted from each of six spectral segments independently.  
The six segments were then combined together to form a spectrum covering  
74 - 111 GHz.  The complete spectrum of each of the 10 galaxies 
is presented in Figure 1.  Note that due to the problems mentioned earlier, 
in many of the galaxies a small spectral window 
around 92 GHz has no usable data.  

The RSR does not have Doppler tracking ability; thus, the spectra have a 
frequency axis in the local rest frame of the telescope.  Due to changes
in the Earth's motion relative to the sources, observations 
made at different times and on different days have slightly different frequency 
offsets, but these offsets are less than 5 MHz, much smaller than
the 31 MHz resolution of the spectrometer.  
The intensity scales of all spectra shown are {\it T$_A^*$}
(antenna temperature corrected only for atmospheric losses).  The 
main beam efficiency of the 14-m telescope using the RSR has been
measured to vary from 0.63 at 89 GHz to 0.51 at 107 GHz.  The decline
in efficiency at
high frequencies is due to the increased amount of power lost
to the error pattern of the telescope.  The beam 
size also varied with frequency, with a HPBW of approximately 70\arcsec\ at
74 GHz decreasing to 47\arcsec\ at 111 GHz.

Since the galaxies observed vary enormously in distance, the regions probed
by our RSR observations vary from spatial scales of approximately 1 kpc for the 
nearest galaxy to 18 kpc for the most distant galaxy.  However, in many of these
galaxies the molecular emission is centrally concentrated \citep{you95}, so that the
observed emission is likely dominated by gas in the central $\sim$1 Kpc region.  It is 
worthwhile noting that the beam size variations over the RSR bandpass could 
distort line ratios if the emission were extended.  However, if the bulk of the
emission arises from a nearly unresolved source, then the line ratios will not be 
affected.

\section{Results}
\subsection{Line-integrated Intensities}

The spectra shown in Figure 1 display a number of strong lines that are in 
common to almost all of the galaxies observed.  In addition to the strong
lines of $^{13}$CO, HCN, HCO$^+$, HNC and CS, many additional much
weaker lines are also detected.  The spectral scans of Orion KL and 
Sgr B2 \citep{tur89} were used as a guide to compile a list of spectral lines and
line blends that are likely to be detectable in these galaxies and
from this compilation we have identified 
all of the spectral features in the 10 galaxies.  Lines from CO, CS, 
HCN, HCO$^+$, HNC, N$_2$H$^+$, C$_2$H, HNCO, CH$_3$OH, HC$_3$N, C$_3$H$_3$, 
SO and CH$_3$C$_2$H are detected in one or more of these galaxies.  In addition,
at least one of the hydrogen recombination lines was detected in M82, NGC 1068
and NGC 4258.  

One of the most important advantages of obtaining each 3 mm wavelength spectrum 
simultaneously with the RSR is the perfect pointing registration in
all spectral lines.  These galaxies all have complex small-scale structure, so
even small pointing errors can seriously affect line ratios.  In addition,
because the lines were obtained simultaneously with a single instrument, the
relative calibration of the spectral line intensities is very good. 
 
The intensities of the spectral features were determined by fitting
each spectrum with a 
template of spectral lines assuming a common line width and redshift. 
The list of lines was developed from the strong lines seen in the spectral
scans of Orion and Sgr B2 \citep{tur89}, and include all plausible lines, 
even if the chemistry is different than these Galactic sources.  
The benefit of such a simultaneous fit is that the line width and redshift can be 
well constrained by the stronger spectral lines, permitting more accurate
determination of the intensities of the weaker lines. Of course this 
assumes that the emission from the various molecular species are distributed 
similarly so they share a common line width and line velocity.  
Examination of the residuals to the template fits finds them consistent with 
the baseline noise, thus, we believe that these assumptions are satisfied. 
The 33 spectral features that were fit
are listed in Table 2, which include the isotopic lines of 
H$^{13}$CN, H$^{13}$CO$^+$, HN$^{13}$C, and C$^{34}$S. Although these
isotopic lines were not detected in any of the galaxies, they were
included in the simultaneous fits to provide upper
limits on the isotopic emission.  Since hydrogen recombination lines
were possibly detected in three of these galaxies, the 
entire series of hydrogen recombination lines were included in the fits.  
In several cases a single spectral feature was fit to a blend of lines 
(marked blend in Table 2) when the line separation in frequency
was much smaller than the 31 MHz spectral resolution of the RSR.  In most 
cases these were blends of lines from the same molecular species. 

Fits to the 33 spectra features are shown superimposed on the spectra 
in Figure 1.  All features with any significance are well 
fit by this line template.  Features detected at a significance of 3$\sigma$ or 
greater are labeled in Figure 1.  The results of our template fitting procedure
for all of the galaxies are summarized in Table 3.  For features detected at 
a significance of 3$\sigma$ or greater, the table provides the peak intensity and 
the 1$\sigma$ uncertainty on these intensities (in parentheses).  For features 
not detected at the 3$\sigma$ significance level, the table provides the
3$\sigma$ upper limit to the intensity of the spectral transition or blend. 
Only one of the six hydrogen recombination lines was detected in M82,
however, two other recombination lines were just slightly below the detection 
threshold and remaining two recombination lines were present at the 2$\sigma$ level.
Thus, the presence of hydrogen recombination lines is probably secure.
In NGC 1068 and NGC 4258 only the H41$\alpha$ recombination line was detected and
in NGC 4258 this was only one of the three lines detected. In both of these galaxies 
there is no evidence for any of the other recombination
lines at comparable strengths as the H41$\alpha$ line as would be expected, therefore, 
the identification of this feature at 92.035 GHz as a hydrogen
recombinations line is suspect. The only other
plausible identification for this feature is a blend of CH$_3$CN lines from 
the J = 5-4 transition, however if correct, CH$_3$CN would be surprisingly
strong in these two galaxies.  

The only spectral line detected in all 10 galaxies is \hco.  \thco\ 
was detected in nine galaxies, HCN, HNC and CS were detected in seven galaxies
and \ceio\ was detected in six galaxies. Because our sample of galaxies
vary in distance from 2.9 to 101 Mpc, the absolute detection threshold 
is not the same for these galaxies, however, in all
subsequent analysis we utilize line ratios.  The
richest spectrum is that for NGC 253, where 16 of the 33 spectral features 
listed in Table 2 were detected.  We note that there have been many 
spectroscopic observations of these same galaxies, and a number 
of the spectral lines in Table 2, not detected in our survey, were 
detected in more sensitive measurements focused on single spectral lines.
In only six galaxies, Arp 220, NGC 1068, NGC 253, M82, IC 342, and Maffei 2,
five or more spectral features are detected.
In NGC 6240 the only line detectable is \hco, and in two other galaxies, 
NGC 3690 and NGC 4258, we detect \hco\ but not HCN.  
In the following section the template fits are used to examine several 
line intensity ratios and investigate variations in these ratios from galaxy 
to galaxy.

\subsection{Measured Line Ratios}
\subsubsection{Isotopic Ratios}

We first examine the CO isotopic ratios.  In the six galaxies where both 
\thco\ and \ceio\ were detected, the average ratio \thco/\ceio\ was 
4.5.  The spread in this ratio was relatively small, ranging from 3.3 in NGC 3079 
to 7.8 in Maffei 2. These measured ratios are similar to 
those found for massive GMCs in the disk of the Milky Way 
\citep{pen72} and for Sgr B2 \citep{cum86}.
A survey of luminous star-forming regions in the Milky Way
\citep{loe09} found \thco/\ceio\ ratios between 3 and 16, with 
the average region having a ratio of about 6, again very similar to the values 
measured in our galaxy sample.  This ratio can be affected by optical depth; 
however, this lowest transition of CO is intrinsically weak, so it is expected 
that these isotopic lines are optically thin.  
Thus, this ratio likely reflects the abundance ratio of the
isotopes of atomic carbon and oxygen.  In the Milky Way 
both the $^{12}$C/$^{13}$C ratio and the 
$^{16}$O/$^{18}$O ratio have a strong dependence on galactic radius \citep{wil99},
both increasing with increasing galactic radius.   
\citet{wil94} suggest that both $^{13}$C and $^{18}$O are secondary
nuclear products, while $^{12}$C and $^{16}$O are primary nuclear products.
Ratios of primary to secondary products, such as $^{12}$C/$^{13}$C, are 
expected to be a good indicator of the chemical evolution of the Galaxy.
However ratios of secondary products, such as \thco/\ceio, should not
vary greatly as is observed.

We did not detect the J=1-0 transitions of H$^{13}$CN, H$^{13}$CO$^+$, 
or HN$^{13}$C or the J=2-1 transition of C$^{34}$S in any of the galaxies
in our sample, however,
these limits provide information on the optical depth of the 
main lines.  The most sensitive limits on the isotopic line strengths 
were set in NGC 253 and M82.  In NGC 253 the 3-$\sigma$ lower limits 
on the isotopic intensity ratios were 
HCN/H$^{13}$CN $>$ 19, HCO$^+$/H$^{13}$CO$^+$ $>$ 16, and CS/C$^{34}$S $>$ 8.
In M82 the 3-$\sigma$ lower limits were HCN/H$^{13}$CN $>$ 12, 
HCO$^+$/H$^{13}$CO$^+$ $>$ 19, and CS/C$^{34}$S $>$ 5.
If we assume isotopic abundances measured or NGC 253
summarized by \citet{omo07}, we conclude that the main isotopic lines of
HCN, HCO$^+$, HNC and CS have optical depths $<$ 3.  

\subsubsection{{\rm HCN/\thco} Ratio}

Both CO and HCN emission are often used as tracers of the molecular gas
in galaxies \citep{omo07}.  Since the transitions of CO and HCN have 
quite disparate critical densities, they may be expected to trace different
density molecular gas, and thus their ratio is a measure of the dense
gas fraction \citep{pag98}.  $^{12}$CO is outside the frequency range of the RSR for these 
very low redshift galaxies; however, we believe that \thco\ may be a much 
better probe of the gas.  The \thco\ emission likely arises from higher 
column density molecular gas than CO, and much of the \thco\ emission 
may be produced in the same gas responsible for the HCN emission.  Thus, 
the HCN/\thco\ ratio may be a good diagnostic 
of the density of the molecular gas most directly connected to the star 
formation activity in the nuclear starburst regions of these
galaxies.  We find that the 
HCN/\thco\ ratio varies significantly
from galaxy to galaxy, with the largest ratio of 2.2 found in Arp 220 and the
smallest ratio found in NGC 4258, where the 3-$\sigma$ upper limit was only 0.30.  
We discuss later how this ratio can provide an important constraint on the mean
gas density.

\subsubsection{{\rm HCO$^+$/HCN} Ratio}

Early models of the chemistry of X-ray irradiated gas suggested that
HCO$^+$ would be under-abundant in gas near a hard X-ray
source \citep{lep96, mal96}.  These results
led to the suggestion that the HCO$^+$/HCN ratio may distinguish 
between PDR and XDR dominated gas and thus between central regions of 
galaxies dominated by either starbursts or AGNs.  Subsequent
observations \citep{koh01, gra06, ima07, baa08} seemed to confirm this prediction.  
These observations also showed an anti-correlation between the observed 
HCN/CO ratio and the HCO$^+$/HCN ratio.  However, the more recent theoretical 
work of \citet{mei05} and \citet{mei07} places this interpretation in question.  
\citet{mei07} showed that in the high density gas, 
the HCO$^+$/HCN ratio is larger in XDR regions than in PDR regions, just 
the opposite of that suggested by \citet{lep96}.  They also noted a strong 
density dependence for this ratio in both PDR and XDR gas. 

In Figure 2 we plot the HCO$^+$/HCN ratio versus the 
HCN/\thco\ ratio.  We find a similar trend to that found in previous observational
papers, that is, the HCO$^+$/HCN ratio is inversely correlated with the
HCN/\thco\ ratio.  The agreement of our result with past papers is not surprising, 
since there is considerable overlap between our galaxy 
sample and those in the previous studies.  We will address the role density
plays in these ratio trends later, however, NGC 3690, NGC 4258 and NGC 6240 
are quite anomalous in having unusually large ratios 
of HCO$^+$/HCN, and variations in chemistry are almost certainly needed to
explain these ratios.  NGC 6240 is not plotted in Figure 2 because
we did not detect either HCN or \thco, but the ratio of HCO$^+$/HCN is greater
than 2.3 (3 sigma limit).  NGC 3690 and NGC 6240 were both
part of the \citet{jun09} study of ULIRGS, and these galaxies were among those
with the largest HCO$^+$/HCN ratio.

\subsubsection{{\rm HNC/HCN} Ratio}

The HNC/HCN ratio is a tracer of gas temperature \citep{aal08}.  
It is well known that in cold clouds the column density 
ratio of HNC/HCN can be as large as 3 to 10 \citep{chu84}, while
in warm GMCs, the column density ratio is less than 1, and can be
as small as 0.015  \citep{gol81}. The
strong dependence of this ratio on temperature is a prediction of molecular cloud 
chemistry \citep{sch92}. Since these lines are similarly excited and have similar
{\it A}-coefficients, when they are optically thin the observed emission ratio 
is proportional to the column density ratio.  For the galaxies in our sample
with detections in both lines, the observed ratio of HNC/HCN is
confined to a narrow range of values between 0.43 and 0.69, 
narrower than that the galaxies observed by \citet{aal02}.  The ratios are
consistent with the emission arising from warm molecular gas.

\subsection{Comparison of M82 and NGC 253}

The two galaxies with the strongest emission lines are 
M82 and NGC 253 which enable a comparison of emission
from other molecular species, such as C$_2$H, N$_2$H$^+$, 
HNCO, CH$_3$OH, HC$_3$N, C$_3$H$_2$, SO and CH$_3$C$_2$H, that had only
a limited number of detections in our galaxy sample as a whole.
In both of these galaxies the emissions observed are likely dominated by 
molecular clouds associated with their nuclear starbursts.  However there
are differences, \citet{ala10}
suggested that the starburst in M82 is more evolved and the
chemistry is dominated by PDR regions, while the chemistry in the
younger starburst, NGC 253, is dominated by shocks.

Notable differences between these two galaxies are the detection 
of HNCO, CH$_3$OH, N$_2$H$^+$ and HC$_3$N in NGC 253 but not in M82, 
although these lines would have been readily detected if they had the same 
line strength relative to HCN as they have in NGC 253. 
Both HNCO and CH$_3$OH are thought to have a shock origin, as these 
molecules are believed to form on grain 
surfaces \citep{tid09} and then ejected from the grains
by low velocity shocks \citep{mar08}. Thus, the presence of
these lines in NGC 253 agrees with the idea that shocks are an important 
component to the chemistry in the nuclear regions of this galaxy.
\citet{omo07} also noted that SiO and CH$_3$CN, both thought to have 
a shock origin, are more abundant in NGC 253 than M82. 
We note that HNCO and CH$_3$OH are even stronger relative to
HCN in IC 342 and Maffei 2, both whose chemistry may also be
dominated by shocks \citep{ala10}.
A similar conclusion concerning IC 342 was reached by \citet{meie05}.  
The largest CH$_3$OH/HCN ratio was measured
in NGC 3079, and thus the chemistry in NGC 253, IC 342, Maffei 2 and NGC 3079
may all be strongly influenced by shocks.  
HC$_3$N was also detected in NGC 253 and not in M82.  The only
other galaxy where HC$_3$N was detected was Arp 220, where the
ratio of HC$_3$N/HCN was about three times larger than in NGC 253.  
Based on the strong emission in Arp 220 \citep{aal02}, \citet{aal07} speculated 
that strong HC$_3$N emission might be indicative of young starbursts and this
perhaps may explain the stronger emission in NGC 253 relative to the
older starburst M82.

The other notable difference is the detection of CH$_3$C$_2$H 
in M82 but not in NGC 253, and the 
CH$_3$C$_2$H/HCN ratio measured in M82 is three times 
larger than the 3$\sigma$ upper limit established
in NGC 253. In fact, other hydrocarbon molecules, such as 
C$_2$H, C$_3$H$_2$ and CH$_3$C$_2$H, have strengths relative to HCN much
larger in M82 than in NGC 253.  The enhancement of these hydrocarbon
molecules in M82 is discussed in \citet{fue05} and they suggested that
photon-dominated chemistry was important for the production of
hydrocarbons in the central region of M82.   
Thus the pattern of lines seen in M82 and NGC 253 supports the suggestion
of \citet{ala10} that the chemistry in NGC 253 is influenced by
shocks, while the chemistry in M82 is more PDR dominated.

\subsection{Starburst Versus AGNs}

From our sample of galaxies, we can also examine the differences 
between the starburst and AGN galaxies.  We have already noted that NGC 3690,
NGC 4258 and NGC 6240, all with AGNs, have anomalously large HCO$^+$/HCN ratio.  
We have formed composite spectra of a sample 
of starburst galaxies (NGC 253, Maffei 2, IC 342, M82 and Arp 220) and galaxies
with AGNs (NGC 1068, NGC 3690, NGC 4258 and NGC 6240).  The emission in each
spectra was first normalized to the strength of the HCO$^+$ line (the only line
seen in all the galaxies), then the frequency axis was adjusted by a
factor of 1+{\it z}, and finally the spectra were averaged using equal weights. Equal
weighting was used so that no galaxy dominated the composite spectrum, 
although this meant that the weaker line galaxies added more noise than
the strong line galaxies to the composite spectra.  The
composite spectra were fit using the same procedure described earlier for the individual
galaxies and, as before, a 3$\sigma$ cutoff was used to define line detections.  
Unfortunately, the AGN composite consisted primarily of weak-line galaxies, so fewer
lines were detected in the composite spectrum relative to the starburst composite.
The most interesting ratio is the HCO$^+$/HCN ratio and in the
starburst galaxies this ratio has a value of 0.71$\pm$0.06, while in the AGN galaxies, the
ratio was 1.9$\pm$0.3, significantly larger.  We can also compare the HCN/$^{13}$CO
ratio, which for the starbursts was 1.11$\pm$0.07 and for the AGNs was 0.64$\pm$0.09.
Ratios of HNC/HCN and CS/HCN were similar in the two samples.  

In the AGN galaxies, NGC 1068 has the strongest lines.  Earlier we noted
that none of the observed line ratios distinguishes NGC 1068 from the sample
of starburst galaxies.  The emission observed in NGC 1068 
may be dominated by the starburst ring and not gas closely associated with the 
Seyfert nucleus.  
Removing NGC 1068, we can form a new AGN composite spectrum from the 
remaining three AGN galaxies.  Although HCN was not detected in any of the 
individual galaxies, it is readily detected in their composite spectrum.
In this new composite AGN spectrum, we find ratios 
of HCO$^+$/HCN = 3.4$\pm$0.9 and HCN/$^{13}$CO = 0.45$\pm$0.12.  Not surprising, 
these ratios are even more extreme relative to the starburst sample than the AGN 
composite including NGC 1068.

\section{Molecular Gas Models}
\subsection{Modeled Line Ratios }

The line ratios discussed in the previous section depend on both the physical
(density, column density, and temperature) and chemical (molecular abundance)
properties of the gas.  The information available in our spectra is inadequate
to constrain all the needed gas properties.  We have consequently
modeled the line ratios with a very simplistic approach in which we assume that
the emission arises from a collection of identical molecular cloud cores which are 
characterized by a single temperature and density,
with equal abundances of the various molecular species.  Molecular abundances 
were taken from the well studied Galactic GMC cores M17 
and Ceph A summarized by \citet{ber97}.  We further assume abundance ratios of 
CO/H$_2$ = 1$\times10^{-4}$ and \thco/CO = 2$\times10^{-2}$; the 
resulting molecular abundances relative to H$_2$ are given 
in Table 4.  An E/A ratio of 1 for CH$_3$OH and an ortho/para ratio of 
3 for C$_3$H$_2$ were also assumed.  The assumed molecular
abundances are very similar to those derived by \citet{mar06} for NGC 253
 based on their 2 mm wavelength spectral survey (see comparison in Table 4). 

The emission from an ensemble 
of GMC cores was modeled with the RADEX code \citep{van07} using the Einstein 
{\it A}-coefficients and collision rate coefficients from the Leiden 
Atomic and Molecular Database \citep{sch05}.  To compute the line
intensities, besides the molecular abundances, the gas density, $n$, gas 
temperature, $T$, and the H$_2$ column density per unit line width,
N(H$_2)$/$\Delta$$v$, need to be specified.  
Our standard GMC core model assumes $T$ = 35 K, $n$ = 1$\times$10$^5$ cm$^{-3}$, 
and N(H$_2)$/$\Delta$$v$ = 1$\times$10$^{22}$ cm$^{-2}$ (km s$^{-1}$)$^{-1}$.   
Line widths for GMCs in the Galactic disk are around 3 km s$^{-1}$ \citep{sne84},
leading to a H$_2$ column density of 3$\times$10$^{22}$ cm$^{-2}$.  Molecular 
clouds near the center of our Galaxy, such as Sgr B2, have line widths
around 20 km s$^{-1}$ \citep{cum86}, and for this line width the modeled
H$_2$ column density is 2$\times$10$^{23}$ cm$^{-2}$. 

The line ratios computed from our standard model are compared with the 
observations of NGC 253, the galaxy with the richest spectrum, in Figure 3.  
Since HCN usually has the strongest emission of any of the 'high dipole moment' 
molecules, all intensity ratios are computed relative to HCN.  Multiple lines
were observed for several molecular species and for the line ratio analysis
the 87.9252 GHz line for HNCO, the 96.7414 GHz line for CH$_3$OH and the 99.2999 GHz 
line for SO were used in forming the intensity ratios.  For HC$_3$N, the two 
lines at 81.8815 and 100.0764 GHz, which
were in the least line confused regions of the spectrum, were
summed together.  The line ratios for our standard core model
agree well with those observed for NGC 253.  The largest discrepancies 
between the observed and modeled line ratios is less
than a factor of two.  The observed ratios
in this galaxy are well fit using this simple model.

Our model can also be used to address the sensitivity of the various 
line ratios to density and temperature.  The line ratios were
computed from the standard model for densities of 1$\times$10$^4$ and 
1$\times$10$^6$ cm$^{-3}$ and the
resulting line ratios shown in Figure 3.  For our assumed abundances,
a density of 1$\times$10$^5$ cm$^{-3}$ is the better fit to the observed
line ratios in NGC 253.  Figure 3 illustrates that several of the line
ratios, due to differences in the critical density of the lines
forming the ratio, are highly sensitive to density.  As expected, one
of the values most affected is the HCN/\thco\ ratio; however, in addition,
the line ratios CH$_3$OH/HCN and HNCO/HCN are also density sensitive.
The temperature sensitivity of these ratios can also be explored, and in 
Figure 4 the observed line ratios in NGC 253 are 
compared to those derived from
our standard model with density 1$\times$10$^5$ cm$^{-3}$ and with varying
temperatures.  Temperature has a much 
smaller effect on the line ratios than does density, largely because the
molecular transitions we observed all arise
from similar energy levels above the ground rotational state.  

NGC 253 has been widely studied and there are a number of estimates of the
density of the molecular gas. Based on fitting multiple lines of CS, 
HCN and HCO$^+$ led \citet{mar05} and \citet{knu07} to estimate a
density of approximately 2$\times$10$^5$ cm$^{-3}$, higher than
an earlier estimate based on H$_2$CO emission \citep{hut97}.
Recently, \citet{bay09} fit the CS emission from NGC 253 with a
two-component model consisting of a cool, lower density 
component and a warm, higher density component.  For the two velocity
features in NGC 253, they
found densities for the cooler component of order 2$\times$10$^4$ and for the
warmer component of order 2$\times$10$^6$ cm$^{-3}$.  The density
in our standard model agrees well with the single
density fits of \citet{mar05} and \citet{knu07} and is  
consistent with the average of the components fit by \citet{bay09}.   
Based on this density, the abundances we assumed in our standard model,
including those molecules (HNCO and CH$_3$OH) that may be produced in shocks, 
are consistent with the observations in NGC 253.  A better determination
of the physical properties of the molecular gas is needed to refine
the abundance estimates.  

Our model can be used to examine the anti-correlation between the
HCO$^+$/HCN and \thco/HCN ratios shown in Figure 2.  Overlaid
on the data in Figure 2 is shown a line connecting the results for five
models in which only the density is varied, from 1$\times$10$^4$ 
to 1$\times$10$^6$ cm$^{-3}$.  The comparison between data and models 
suggests that much of the variation observed in the \thco/HCN ratio 
can be explained if the average density of the molecular gas
varies among these galaxies.  For densities
below 1$\times$10$^4$ cm$^{-3}$, it is difficult to produce detectable
HCN or \thco\ emission without having unrealistically large column densities 
of these molecular species.  Although the HCO$^+$/HCN ratio is
not strongly affected by density, some of the anti-correlation found between
this ratio and the \thco/HCN ratio can be explained by variations in the average density.
\citet{jun09} came to similar conclusion 
and the sensitivity of these line ratios to density is also apparent
in the modeling presented by \citet{mei07}.  Based on the line ratios, and
ignoring any chemical differences, the data suggest that the 
molecular gas in Arp 220 is the densest and the molecular gas in
NGC 3079 the least dense.  However, density alone cannot explain the large 
HCO$^+$/HCN line ratios measured in NGC 3690, NGC 4258 and NGC 6240, all
with AGNs.  In the models of \citet{mei07}, only the XDR models
predict HCO$^+$/HCN ratios greater than one, thus, if these
XDR models are correct, it is possible that the AGN has an
influence on the chemistry and is responsible for the
enhanced HCO$^+$ abundance in the central regions of these three galaxies. 
However, the large sample of ULIRGS presented by \citet{jun09} show no
correlation between the HCO$^+$/HCN ratio and whether these galaxies
are believed to be AGN or star formation dominated, putting some doubt in
this AGN hypothesis for the enhanced HCO$^+$ emission.

The comparison between models and observations can
be extended to the line ratios measured for 
IC342, M82, Arp 220, Maffei 2, NGC 1068 and NGC 3079.  These
comparisons are shown in Figure 4.  Since density has the greatest
influence on the line ratios, only models with varying density at
a fixed temperature of 35 K are included in this figure.
Based on model comparisons shown in Figures 2 and 4, 
IC 342, Maffei 2 and NGC 3079 would be 
expected to have the lowest average gas density of $\sim$10$^4$ cm$^{-3}$.
This low average density is borne out by the other density sensitive 
ratios.  With this density, we find the abundances from our 
standard core model reproduce well the measured line ratios in these three
galaxies. Similarly, the average density for M82 
is expected to be somewhat larger ($\sim$3$\times$10$^4$ cm$^{-3}$) and
assuming this density we confirm that the abundances of 
HC$_3$N, CH$_3$OH, N$_2$H$^+$, and HNCO relative to HCN are smaller 
than in our standard model and for those found in NGC 253. 
Again based on the result shown in Figures 2 and 4,
NGC 1068 is expected to have a somewhat larger average density
($\sim$10$^5$ cm$^{-3}$), while Arp 220 is expected
to have the highest average density ($\sim$10$^6$ cm$^{-3}$).
As suggested earlier, we find an overabundance of 
HC$_3$N relative to HCN in Arp 220 for any of the modeled densities.  
Thus if we allow for 
the average molecular gas density to vary
from galaxy to galaxy, our standard model with fixed abundances 
provides a reasonable good fit to the measured ratios in these galaxies.  

There have been a number of estimates of the average molecular gas density
for galaxies in our sample.  Most of these estimates utilize multiple 
transitions of a single molecular species, thus they are independent
of derived abundances.   However, despite this, estimates for a single galaxy
can still differ by order of magnitude.  Density estimates for
IC 342 \citep{hut97,mei05}, Maffei 2 \citep{hut97}, M82 \citep{nay10}, 
and Arp 220 \citep{gre09} are in reasonable
agreement with our estimates.  The HCN line SEDS shown in \citet{knu07} are
also consistent with the trend in density variations for the galaxies M82, NGC 253 
and Arp 220.  Fitting the gas in the central regions of these galaxies with a 
single density is an oversimplification and \citet{bay09} and \citet{ala10}
have fit multiple-density components for NGC 253, M82, IC 342, Maffei 2 and
NGC 1068.  With the exception of M82, their column density weighted average
density is consistent with our findings.  For M82, both find a column density
weighted density about an order of magnitude higher than that of \citet{nay10}.

\subsection{Molecular Mass and Beam Filling Factor}

The simple model we presented in the previous section, in which the molecular 
emission observed arises from
an ensemble of identical cloud cores, can be used to 
estimate the beam filling factor and mass of molecular gas, 
independent of the {\it X}$_{CO}$-factor.  Since the distances 
of the galaxies in our sample vary
enormously, the region probed ranges from just the central starburst
core to nearly the entire galaxy.  The linear
diameter of the FCRAO beam at the frequency of the \thco\ line for the 
distance of these galaxies ({\it D}$_B$) is given in Table 5.  

The beam surface area filling factor ({\it f}$_{area}$) is related
to the observed antenna temperature of the line corrected for the
main beam efficiency ({\it T}$_{MB}$), the line intensity from our model
({\it T}$_{model}$), the assumed core line width in the model 
($\Delta v_{model}$), and the observed line width of the galaxy 
($\Delta v_{obs}$), which is dominated by galactic rotation,
by the following equation:

\begin{equation}
f_{area} = \frac{T_{MB}}{T_{model}} \frac{\Delta v_{obs}}{\Delta v_{model}}.
\end{equation}

\noindent
The observed \thco\ emission is used to estimate the beam filling factor
from the above relation.  The modeled intensity ratios,
discussed previously, did not depend on the line width, however, the total
integrated intensity, and hence the beam surface filling factor, does.  
We have no information
on the line widths of cloud cores in these galaxies, so we assume a line width
of 10 km s$^{-1}$ for our model cores, which corresponds to an H$_2$ column
density of  1$\times10^{23}$ cm$^{-2}$.  Although a density of 
1$\times10^5$ cm$^{-3}$ was assumed, the emission from \thco\ has very little
density dependence as long as {\it n} $\geq$ 1$\times10^4$ cm$^{-3}$.  Based
on the observed line intensities and line
widths presented in Table 3, the computed beam area filling factor is
summarized in Table 5.  Not surprising, the beam filling factor is strongly
dependent on galaxy distance.  For the nearer galaxies, the beam area filling factor 
can be as large as 0.26, implying that one-fourth of the FCRAO 
$\sim$50\arcsec\ beam is covered with
gas with a total gas column density of 1$\times10^{23}$ cm$^{-2}$, 
corresponding to an {\it A}$_v$ of approximately 100 mag.  
Similar beam area filling factors would have been found if we 
had used other molecular species, such as HCN or HCO$^+$.   However, 
the results would have been more sensitive to the assumed density of 
the cloud cores.

The molecular gas mass can also be estimated from our model.  The
mass is given by:

\begin{equation}
M = m_{H_2}\: N_{model}\: f_{area}\: d^2\: \Omega_B,
\end{equation}

\noindent
where {\it m}$_{H_2}$ is the mass of a molecular hydrogen molecule,
{\it N}$_{model}$ is the gas column density in the core model, {\it d} is
the distance to the galaxy, and $\Omega_B$ is the main beam
solid angle.  As long as the \thco\ emission is optically thin, 
the mass determination is 
independent of many of the assumed core model parameters.  However,
two important parameters in deriving masses 
are (1) the assumed core gas temperature, as
this affects the partition function, and (2) the assumed \thco/H$_2$ ratio of 
2$\times10^{-6}$.  The mass of molecular hydrogen derived from
the above equation is summarized in Table 5.
Because of the varying distance of these galaxies, the mass
of the nearby galaxies (NGC 253, Maffei 2, IC 342 and M82) includes
only the gas in the central 1 kpc regions, whereas for the most distant 
galaxies (NGC 3690 and Arp 220) it includes the entire molecular mass of
the galaxy.

These mass estimates can be compared with those derived from CO using
an assumed {\it X}$_{CO}$-factor.  To make certain that we are comparing the mass
from comparable regions of these galaxies, we have used CO observations
also obtained with the FCRAO 14-m telescope
as part of the FCRAO Extragalactic CO Survey \citep{you95}.  Observations
of all of the galaxies except Maffei 2 are covered in this survey, and
these  observations
were obtained with nearly an identical beam size as our observations of
\thco.  The CO fluxes from the survey are summarized in Table 5, where
the integrated intensity to flux conversion factor
provided in \citet{you95} was used.

To estimate the mass from the CO fluxes, an {\it X}$_{CO}$-factor of 
1.8$\times10^{20}$ cm$^{-2}$ (K km s$^{-1}$)$^{-1}$, was used which is
representative of Galactic molecular clouds \citep{omo07}.  The molecular
hydrogen gas mass can then be written \citep{ken89} as:

\begin{equation}
M(H_2) = 7.1\times10^3 d^2 S_{CO},
\end{equation}

\noindent
where d is the distance in Mpc and {\it S}$_{CO}$ is CO flux in units of Jy km s$^{-1}$.
The masses derived from this expression are presented in Table 5.  Mass 
estimates from CO are 1.2 - 4.3 times larger than that derived from \thco.  
It is possible to reconcile
these masses if we either lower the \thco\ abundance, increase the gas
temperature assumed in our core model, or decrease the {\it X}$_{CO}$-factor used.  
We cannot rule out any of these possibilities; however, a more likely
scenario is that some of the CO emission is arising from lower density and lower
column density gas that does not produce significant \thco\ emission.  
No matter what the reason for the mass difference, we can conclude that the 
mass of dense gas must represent a significant fraction of the total molecular 
gas in these galaxy centers.  

We find that between about 25 and 85 \% of the total molecular mass 
in these galaxies must have high density and column density.  These
properties are likely associated with the molecular clouds in the central 
$\sim$1 kpc regions of these galaxies.  
\citet{bal87} estimated 
the mean density of the molecular gas in the inner
500 pc radius region of the Milky Way to be 2$\times10^4$ cm$^{-3}$, similar to
the average density estimated for 
several of the galaxies in our survey.  However, the properties of the
molecular clouds in the disk of the Milky Way are quite different.  
\citet{lee90} mapped a 3 deg$_2$ region around $\ell$ = 24\arcdeg\ in both \thco\
and CS.  Although, no estimate was made of the mean density of the gas, the 
quoted emission ratio of CS/\thco\ of 0.008 is substantially lower than the range
of ratios, 0.12 - 1.0, we measured for our galaxy sample.  The largest CS/\thco\
ratio was observed in Arp 220, which is also modeled to have the highest mean
molecular gas density.  Since this ratio is sensitive to the mean 
gas density, this comparison suggests that a large fraction of the molecular gas in 
the central regions of the galaxies in our sample, and the Milky Way, 
must be dense, while molecular clouds in the Milky Way disk have a much smaller
fraction of dense gas.  Studies of individual molecular clouds in the disk of
the Milky Way \citep{car95} confirm this and show that high density and high column 
density cores represent only a small fraction of the area and mass of the 
molecular cloud.

\section{Summary}

We have presented the 3 mm wavelength spectra of the central 
regions of ten relatively nearby galaxies.  The spectra of each galaxy
were obtained simultaneously with the RSR.  
The important advantages of obtaining 
the 3 mm wavelength spectra simultaneously are the perfect pointing registration in
all spectral lines and the very good relative calibration of the spectral line 
intensities.  The only uncertainty in relative 
line intensities comes from the small variations of beam efficiency and beam
size with frequency.  Most of the lines that we detected arise
from high-dipole moment molecules, which allows us to examine the physical
and chemical properties of the dense gas in these galaxies.
Although the galaxies in our survey are far from
being a homogeneous sample, their differences can be exploited to
study the extent to which various line ratios are found to vary and to
what extent they require differences in the physical or chemical properties of
the molecular gas.  

The spectra were fit with a template with 33 spectral features to determine
the line intensities.  We detected 20 of the 33 spectral features in 
our template, and these features arose from 14 different atomic 
and molecular species.  Based on the fitted line intensities, a number of 
line intensity ratios were examined.  The \thco/\ceio\ and HNC/HCN line
ratios are insensitive to density.  The \thco/\ceio\ ratio was found to 
have a very narrow range of values, from 3.3 to 7.8, supporting the idea
that both $^{13}$C and $^{18}$O are secondary nucleosynthesis products.  
The HNC/HCN ratio also was relatively constant from galaxy to galaxy, 
varying between 0.4 - 0.7 and consistent with the emission arising in relatively 
warm gas. 

The emission from these galaxies was modeled with an ensemble of molecular cloud
cores, each with a temperature of 35 K and a molecular hydrogen column density 
per unit line width of 1$\times10^{22}$ cm$^{-2}$ (km s$^{-1}$)$^{-1}$. 
The molecular abundances were taken from well-studied Galactic clouds, although
these abundances are consistent with those found in NGC 253 \citep{mar06}.
The sensitivity of various line ratios to 
density and temperature was examined and it was found that density has
the greatest affect.  The HCN/\thco\ ratio is particularly density sensitive, 
and we believe that the large variation observed for this ratio may be due to differences 
in the mean gas density in these galaxies.  
Varying the density from 10$^4$ to 10$^6$ cm$^{-3}$
can explain the range of HCN/\thco\ ratios observed and provide consistent
results with the other density sensitive line intensity ratios.  Varying the density
can also help explain much of the observed variations in the HCO$^+$/HCN ratio.  
Thus, a simple model with fixed molecular abundances and varying density, 
fits well the 3 mm spectra of most of these galaxies.  
One of the most notable exceptions is the anomalously large HCO$^+$/HCN ratios 
detected in NGC 3690, NGC 4258 and NGC 6240, which cannot be fit without 
increasing the abundance of HCO$^+$ relative to HCN relative.  Based on
the XDR model of \citet{mei07}, a plausible
explanation is that the AGN activity
in these galaxies is responsible altering the chemistry in these 
galaxies and producing the large HCO$^+$/HCN ratio. 

The cloud core model was used to estimate the filling factor of dense, high column
density gas in our beam and to estimate the mass of dense molecular gas.  The 
derived beam filling factors are strongly correlated with distance, as our beam 
varies in linear size from about 0.8 Kpc in the nearest galaxies to over 17 Kpc in
the most distant galaxies.  In the nearby galaxies, beam filling factors as great as 0.26
are found, suggesting that the central regions of these galaxies have
a high filling factor of dense, high column density gas.  We compared our mass of 
dense gas with estimates based on CO and an assumed
{\it X}$_{CO}$-factor and find that the dense gas represents approximately 25-85 \% of the total
molecular gas in the central regions of these galaxies.

\acknowledgements

This work was supported by NSF grants AST 0096854, AST 0704966 and AST 0838222 and by
NASA grant NNG04G155A.  We thank Hugh Crowl and Bing Jiang who assisted in data
collection and Lauren Harley who assisted
in some of the data reduction and spectral line fitting.

\clearpage
\begin{figure}
\begin{center}
\figurenum{1a}
\includegraphics[angle=-0,scale=1]{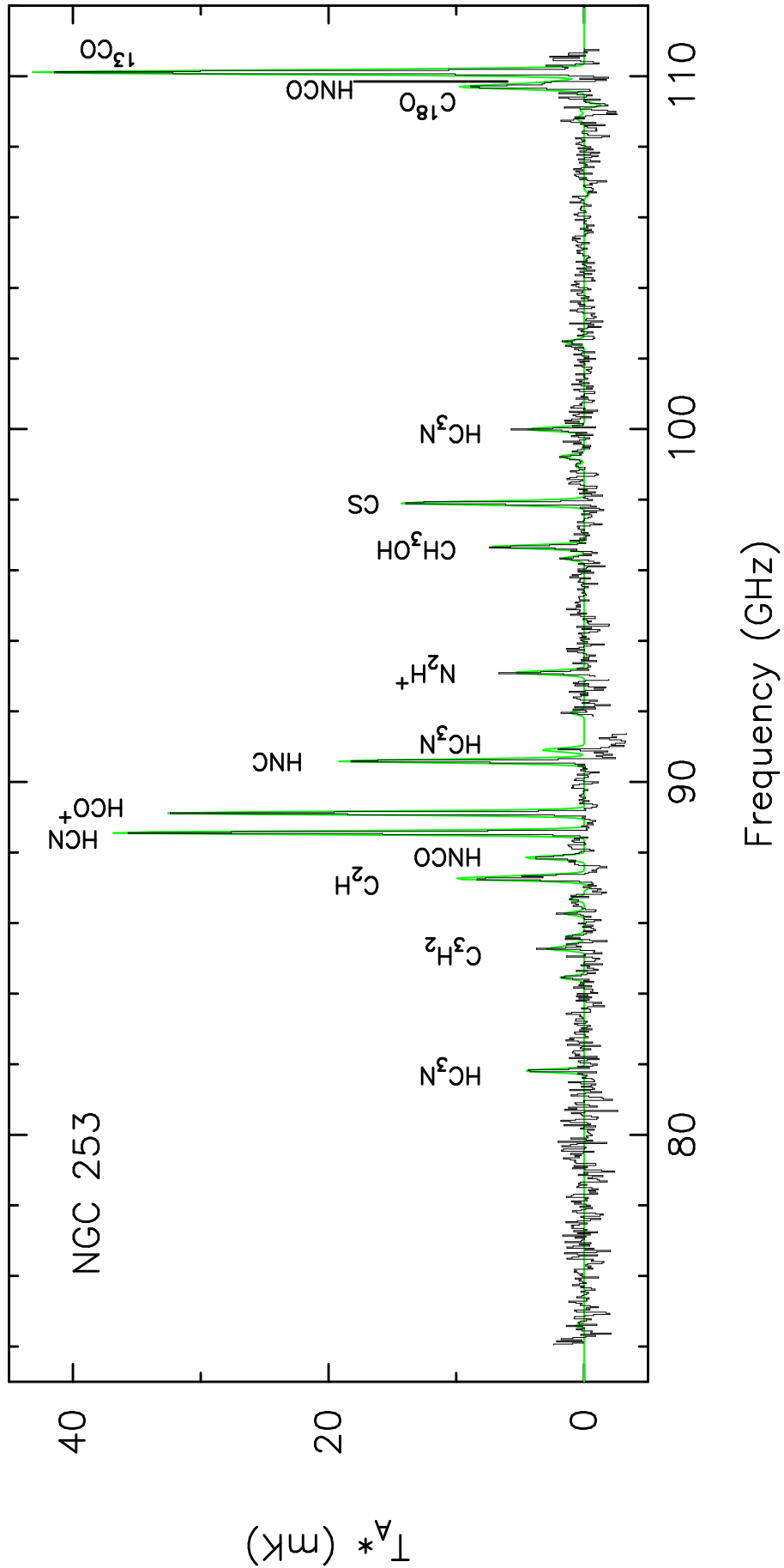}
\figcaption{Three-millimeter spectra for NGC 253 are presented by the binned
points.  The frequency axis is measured relative to the local rest frame of 
the telescope with no corrections for the Earth's motion.  The continuous line (continuous 
green line in the electronic version of the paper)
shows the fit of our 33-spectral feature template discussed 
in the text.}
\end{center}
\end{figure}

\clearpage
\begin{figure}
\begin{center}
\figurenum{1b}
\includegraphics[angle=-0,scale=1]{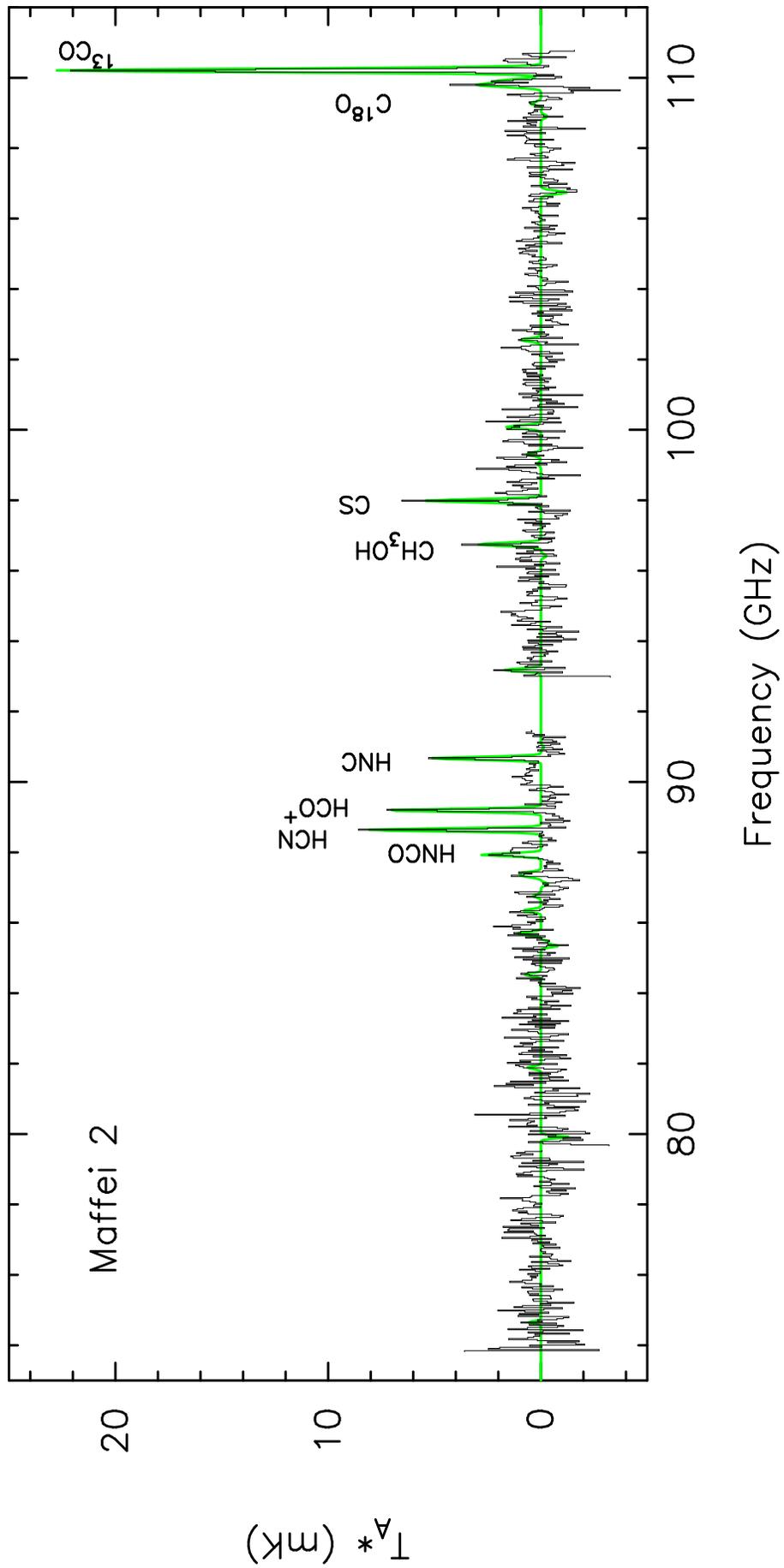}
\figcaption{Same as (a) for Maffei 2. }
\end{center}
\end{figure}

\clearpage
\begin{figure}
\begin{center}
\figurenum{1c}
\includegraphics[angle=-0,scale=1]{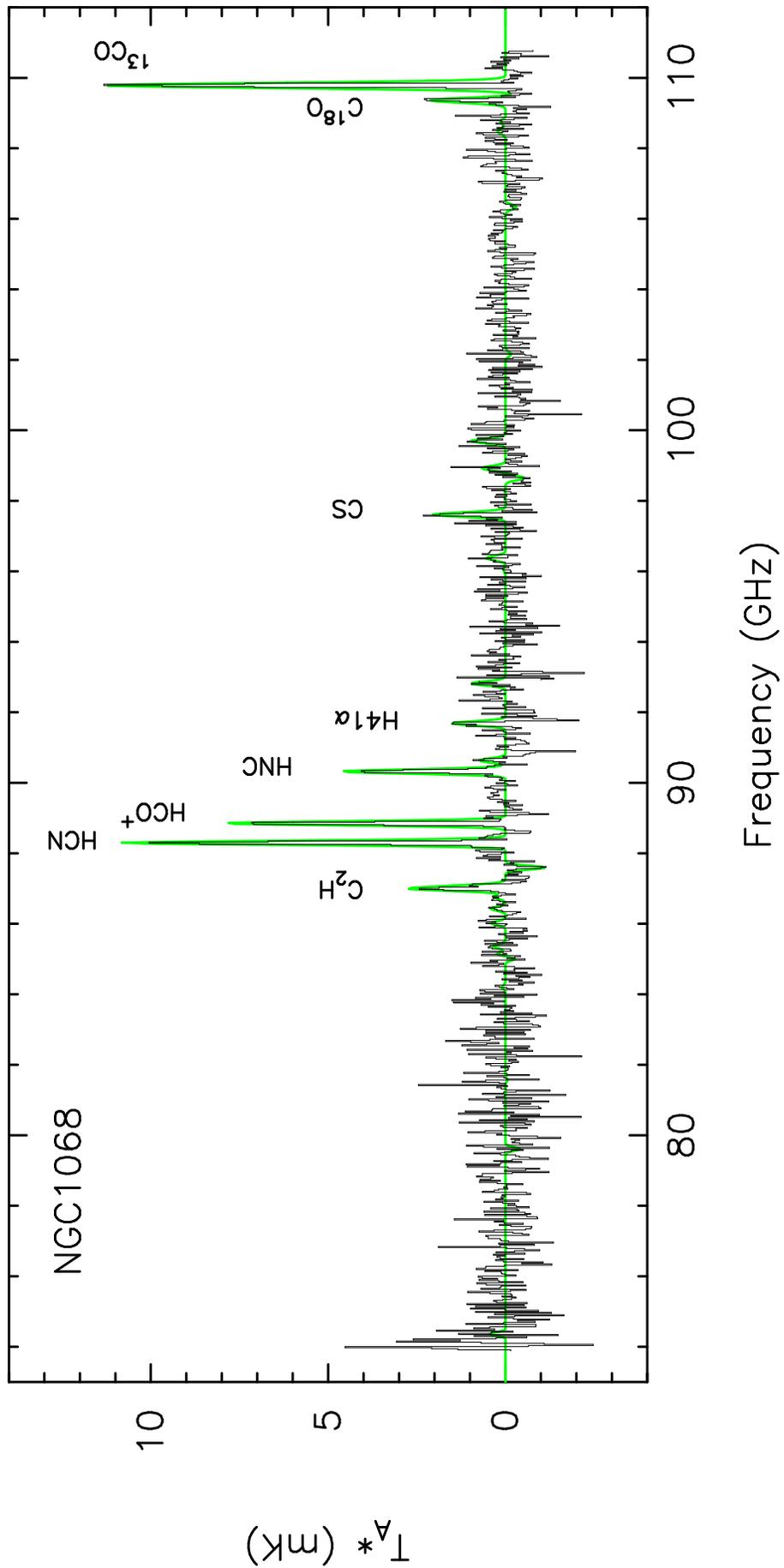}
\figcaption{Same as (a) for NGC 1068.}
\end{center}
\end{figure}

\clearpage
\begin{figure}
\begin{center}
\figurenum{1d}
\includegraphics[angle=-0,scale=1]{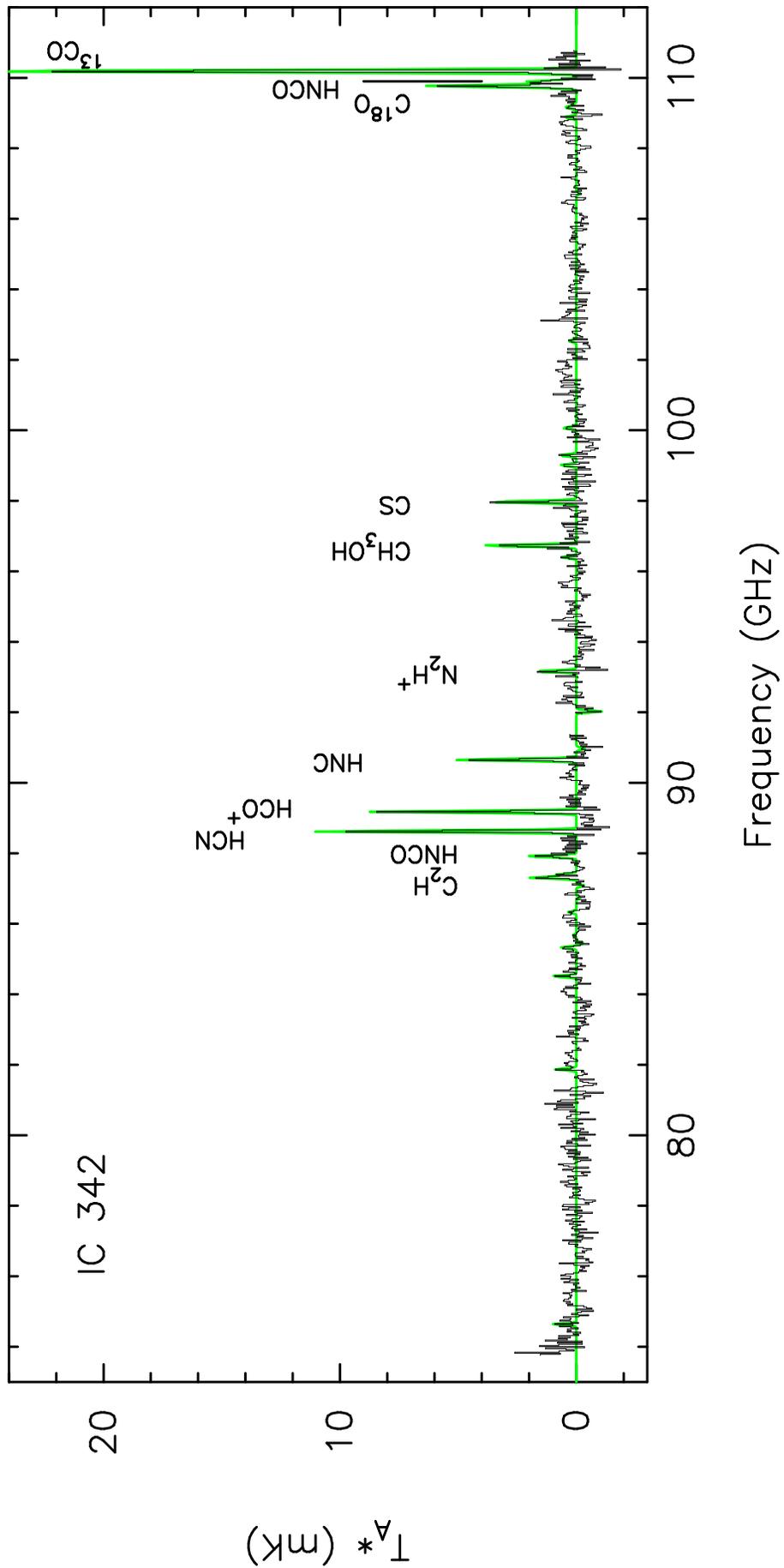}
\figcaption{Same as (a) for IC 342.}
\end{center}
\end{figure}

\clearpage
\begin{figure}
\begin{center}
\figurenum{1e}
\includegraphics[angle=-0,scale=1]{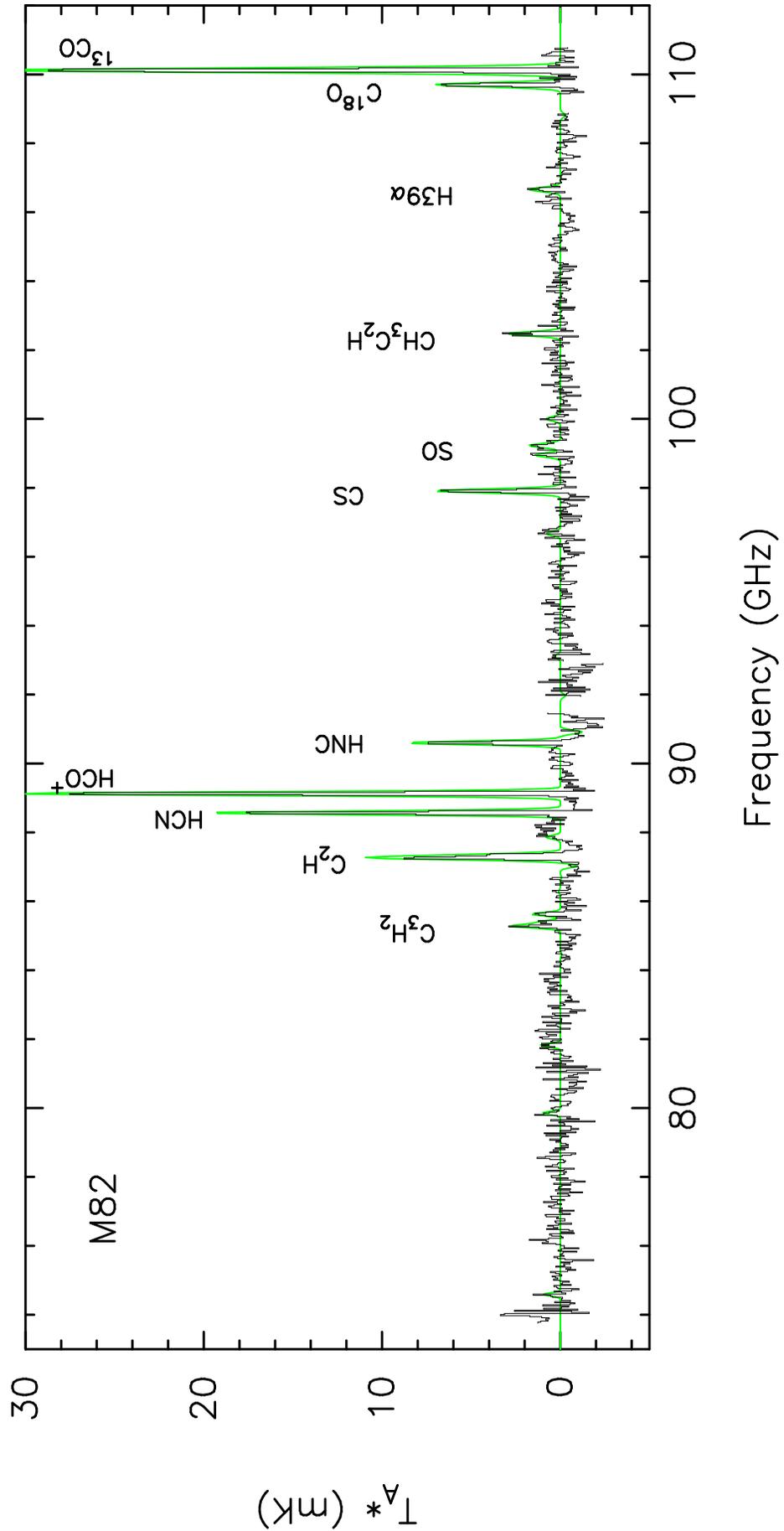}
\figcaption{Same as (a) for M82.}
\end{center}
\end{figure}

\clearpage
\begin{figure}
\begin{center}
\figurenum{1f}
\includegraphics[angle=-0,scale=1]{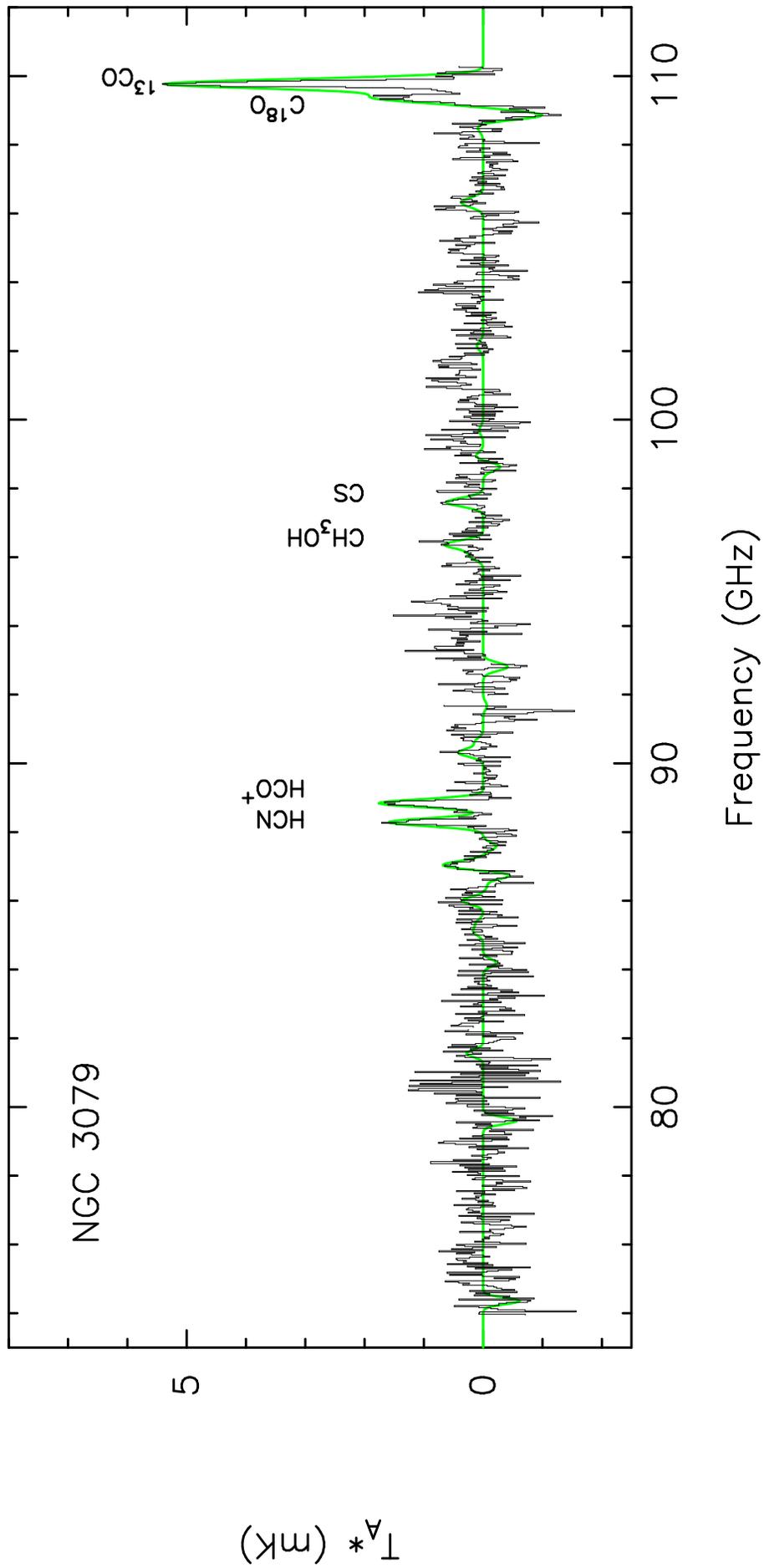}
\figcaption{Same as (a) for NGC 3079.}
\end{center}
\end{figure}

\clearpage
\begin{figure}
\begin{center}
\figurenum{1g}
\includegraphics[angle=-0,scale=1]{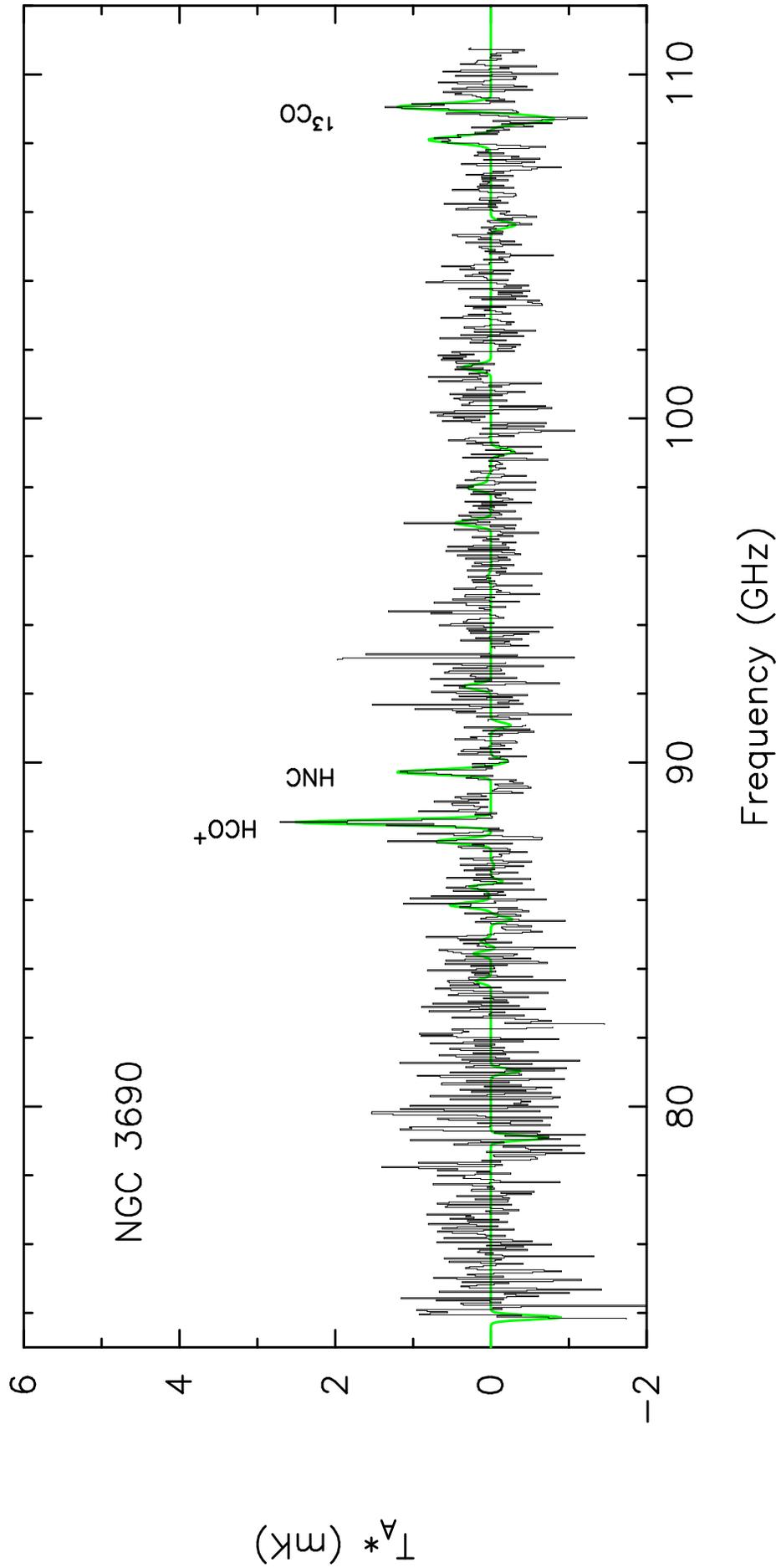}
\figcaption{Same as (a) for NGC 3690.}
\end{center}
\end{figure}

\clearpage
\begin{figure}
\begin{center}
\figurenum{1h}
\includegraphics[angle=-0,scale=1]{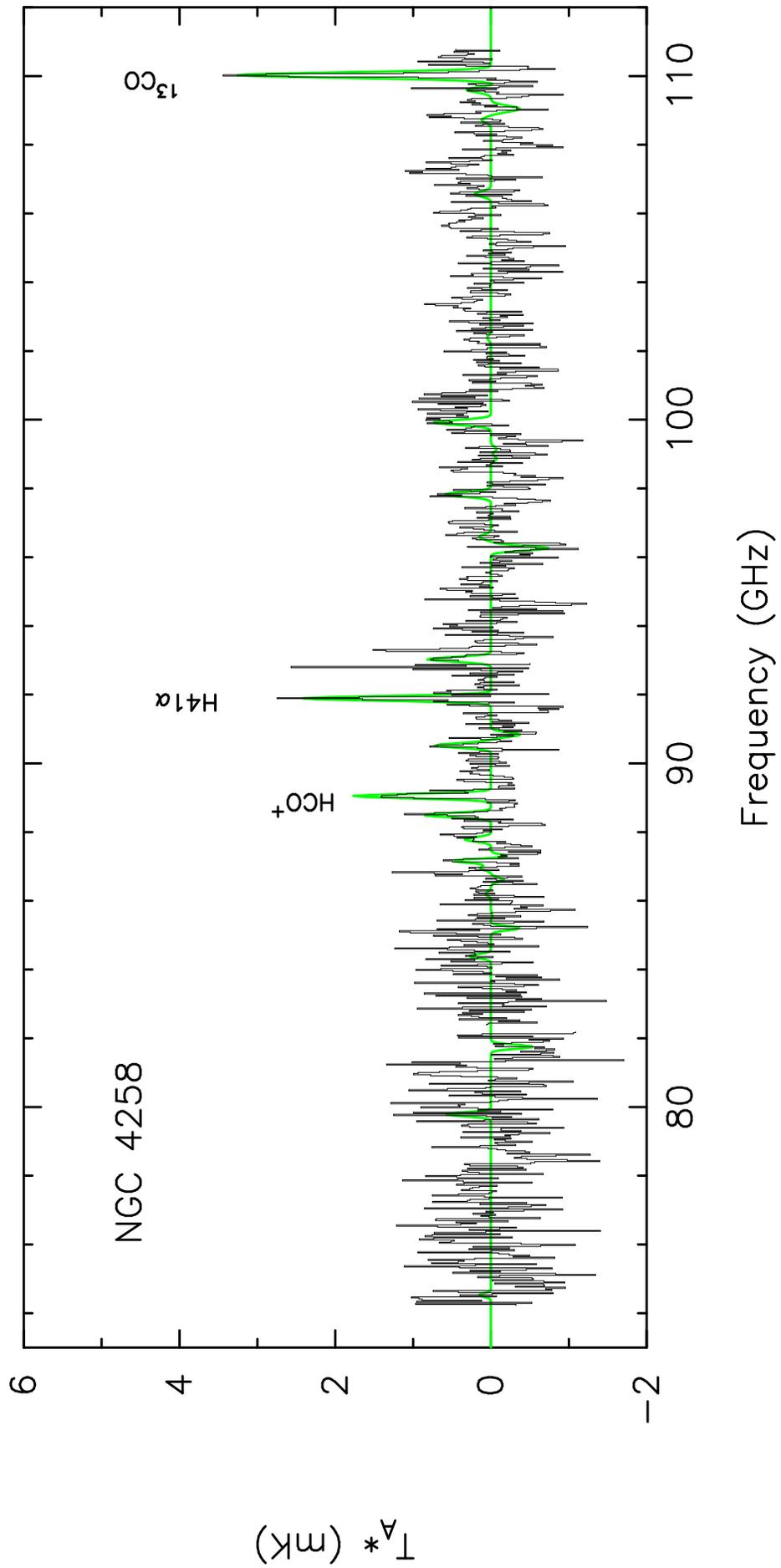}
\figcaption{Same as (a) for NGC 4258. }
\end{center}
\end{figure}

\clearpage
\begin{figure}
\begin{center}
\figurenum{1i}
\includegraphics[angle=-0,scale=1]{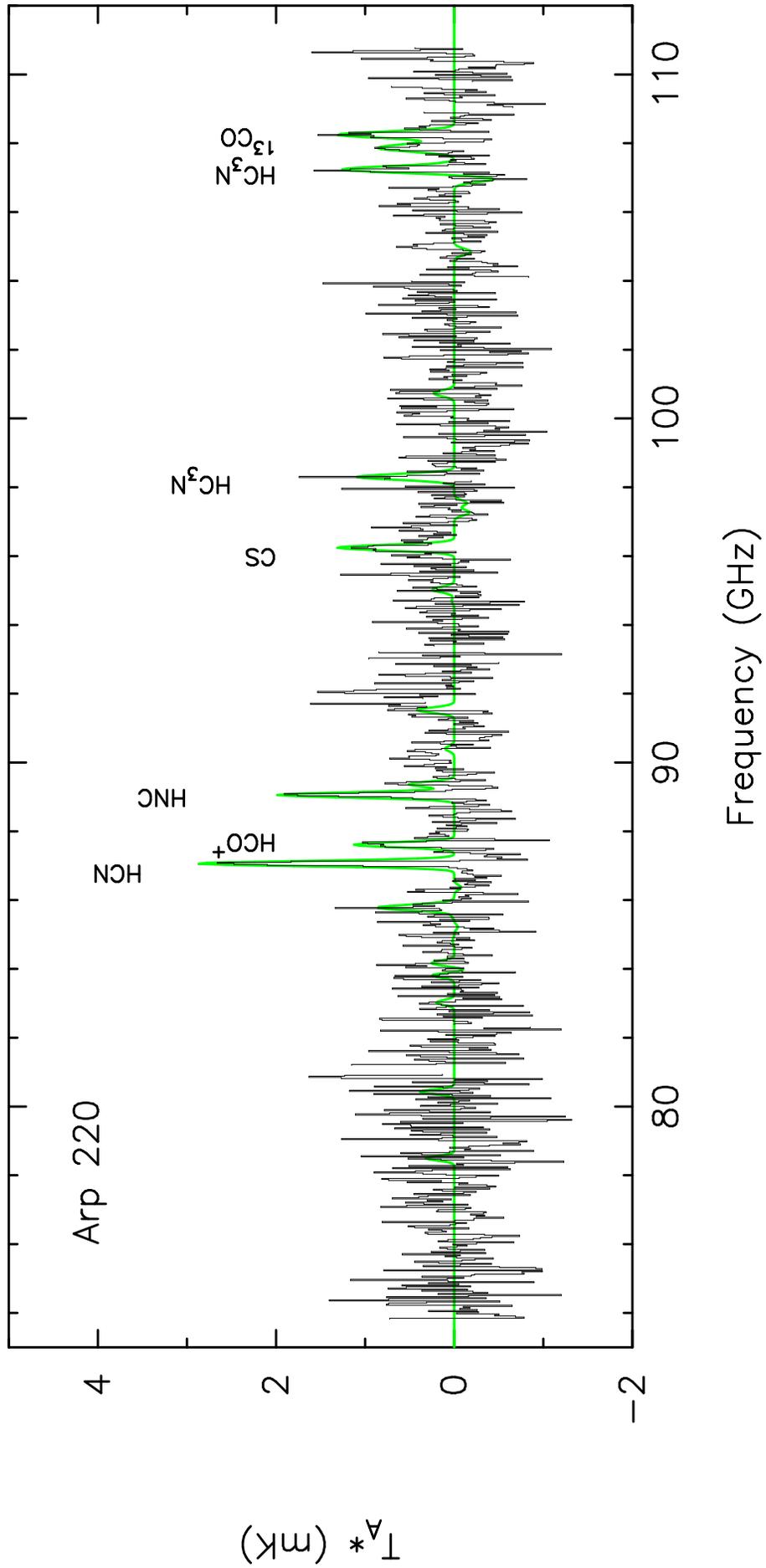}
\figcaption{Same as (a) for Arp 220.}
\end{center}
\end{figure}

\clearpage
\begin{figure}
\begin{center}
\figurenum{1j}
\includegraphics[angle=-0,scale=1]{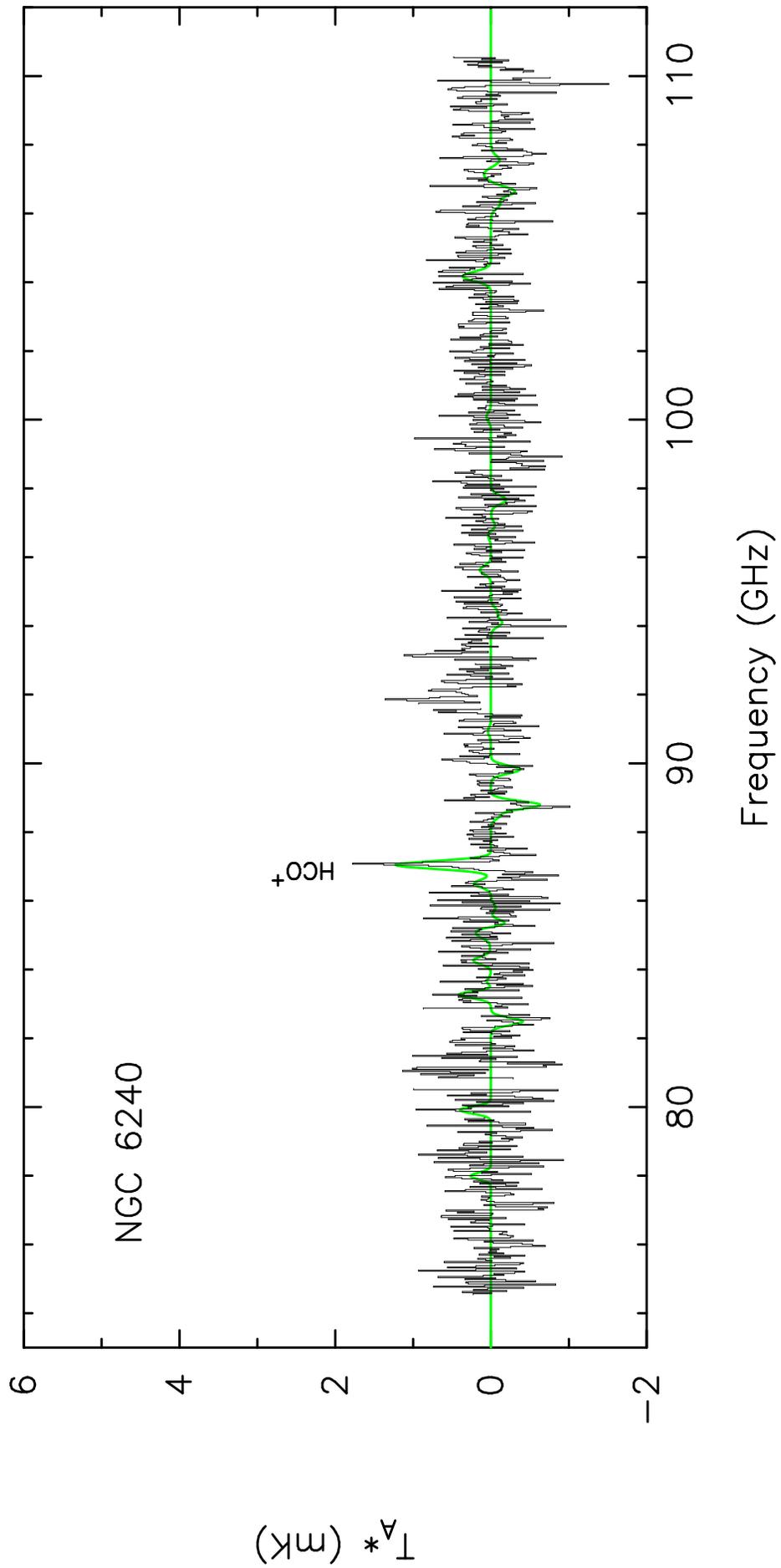}
\figcaption{Same as (a) for NGC 6240.}
\end{center}
\end{figure}

\clearpage
\begin{figure}
\begin{center}
\figurenum{2}
\includegraphics[angle=-90,scale=0.8]{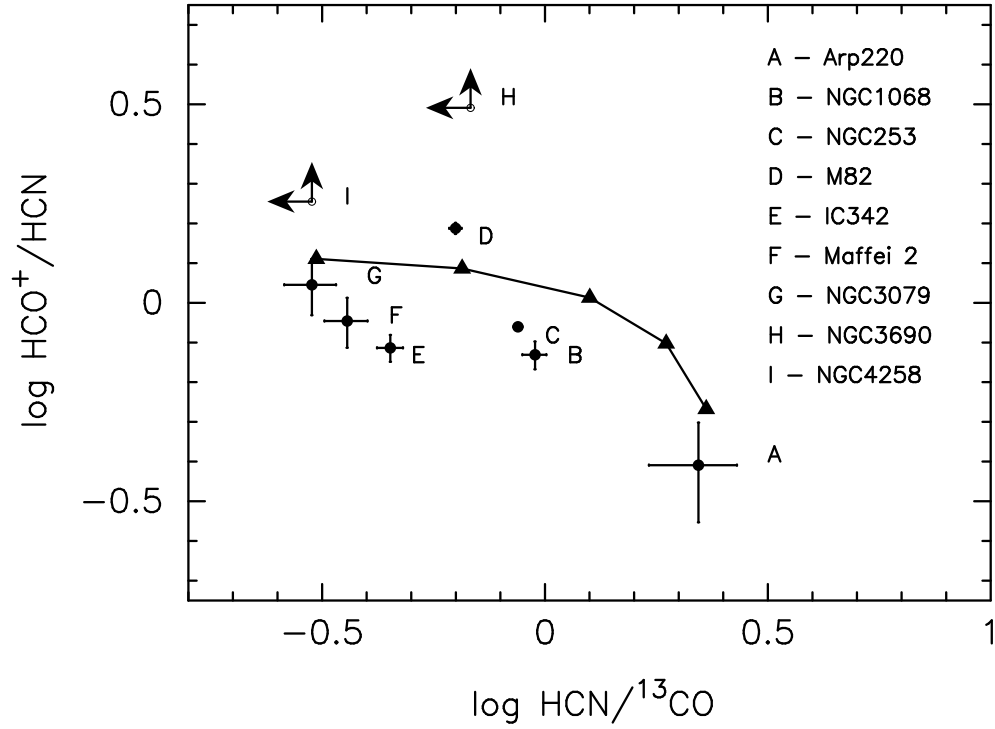}
\figcaption{A plot of the observed intensity ratio of HCN/\thco\ vs. the observed
intensity ratio of HCO$^+$/HCN for the galaxies (filled circles) where \thco\ was detected.  
The error bars show the 1$\sigma$ uncertainty in the values of these ratios.  In the two galaxies
where HCN was not detected, the open circles with arrows mark the 3$\sigma$ upper limit to the
HCN/\thco\ ratio and the 3$\sigma$ lower limit to the HCO$^+$/HCN ratio.  The results of our 
modeling of line ratios (see the text in Section 4.1) are shown by the filled triangles with a 
line connecting five model results with varying density.  The models, from left to right,
have density of log {\it n} = 4.0, 4.5, 5.0, 5.5 and 6.0.}
\end{center}
\end{figure}

\clearpage
\begin{figure}
\begin{center}
\figurenum{3a}
\includegraphics[angle=-90,scale=0.8]{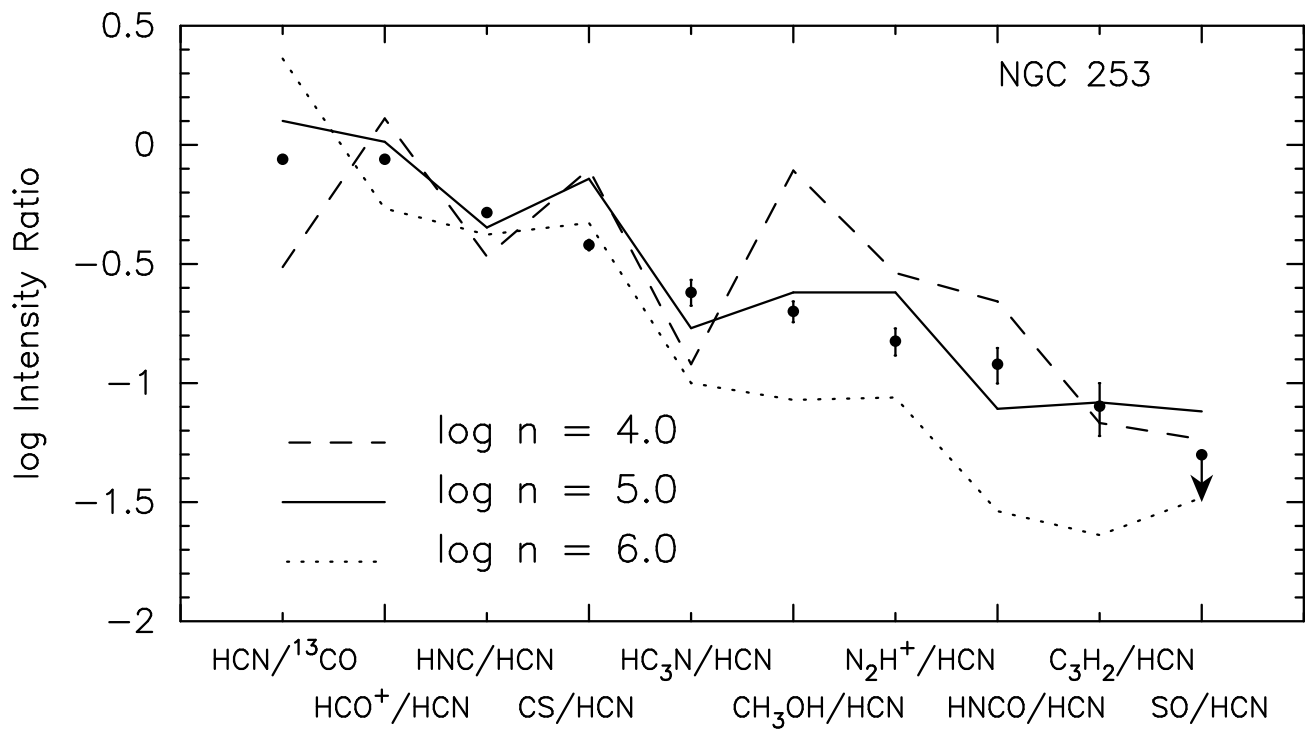}
\figcaption{The observed intensity ratios (shown as filled circles) for a number of key 
molecular species is shown for NGC 253.  The error bars show the 1$\sigma$ uncertainty in 
the values of the line ratios.  Where the molecular species was not detected, the 3$\sigma$
upper limit on the ratio is shown.  Accompanying the observed intensity ratios, are ratios 
computed from our standard core model, see the text; the modeled line intensity ratios are connected by
a dashed line for log {\it n} = 4.0, a solid line for log {\it n} = 5.0 and a
dotted line for log {\it n} = 6.0. }
\end{center}
\end{figure}

\clearpage
\begin{figure}
\begin{center}
\figurenum{3b}
\includegraphics[angle=-90,scale=0.8]{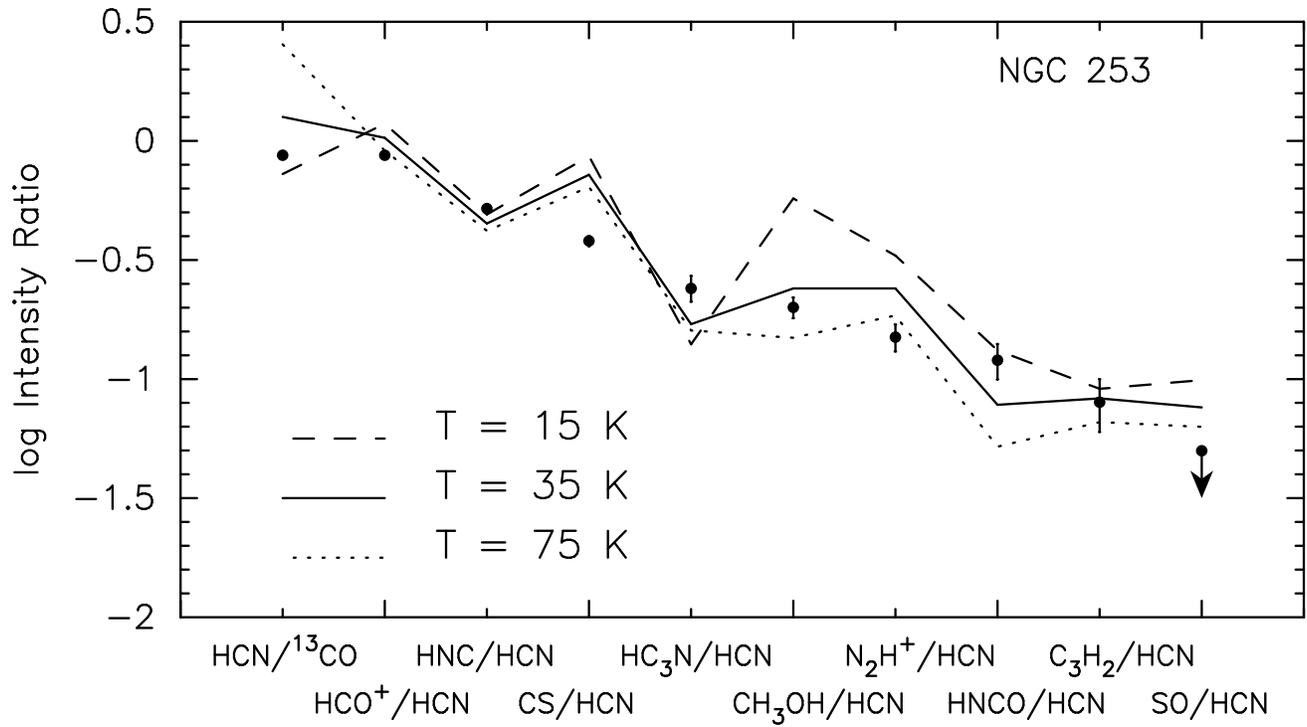}
\figcaption{Same as (a), except showing models of
the line intensity ratios for fixed density of log n = 5.0, and temperatures of 
15 K (dashed line), 35 K (solid line), and 75 K (dotted line).}
\end{center}
\end{figure}

\clearpage
\begin{figure}
\begin{center}
\figurenum{4a}
\includegraphics[angle=-90,scale=0.8]{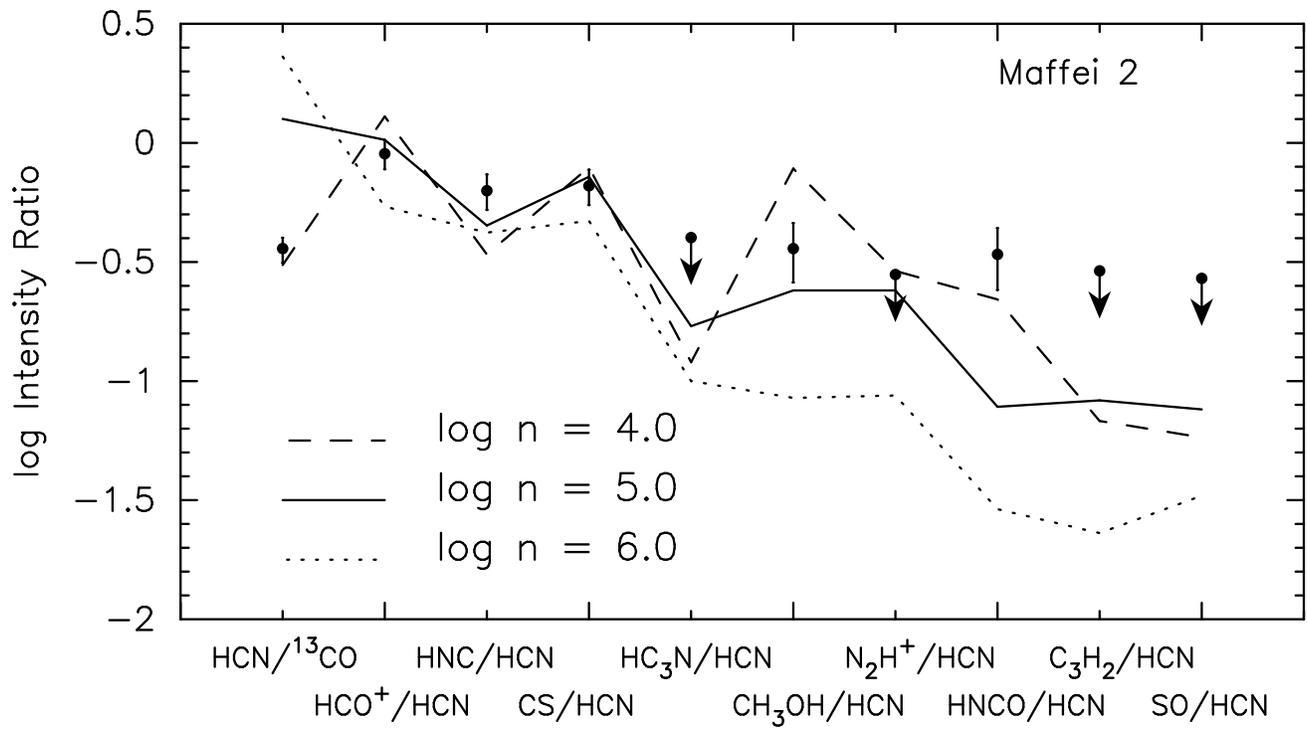}
\figcaption{Same as Figure 3(a) for Maffei 2.}
\end{center}
\end{figure}

\clearpage
\begin{figure}
\begin{center}
\figurenum{4b}
\includegraphics[angle=-90,scale=0.8]{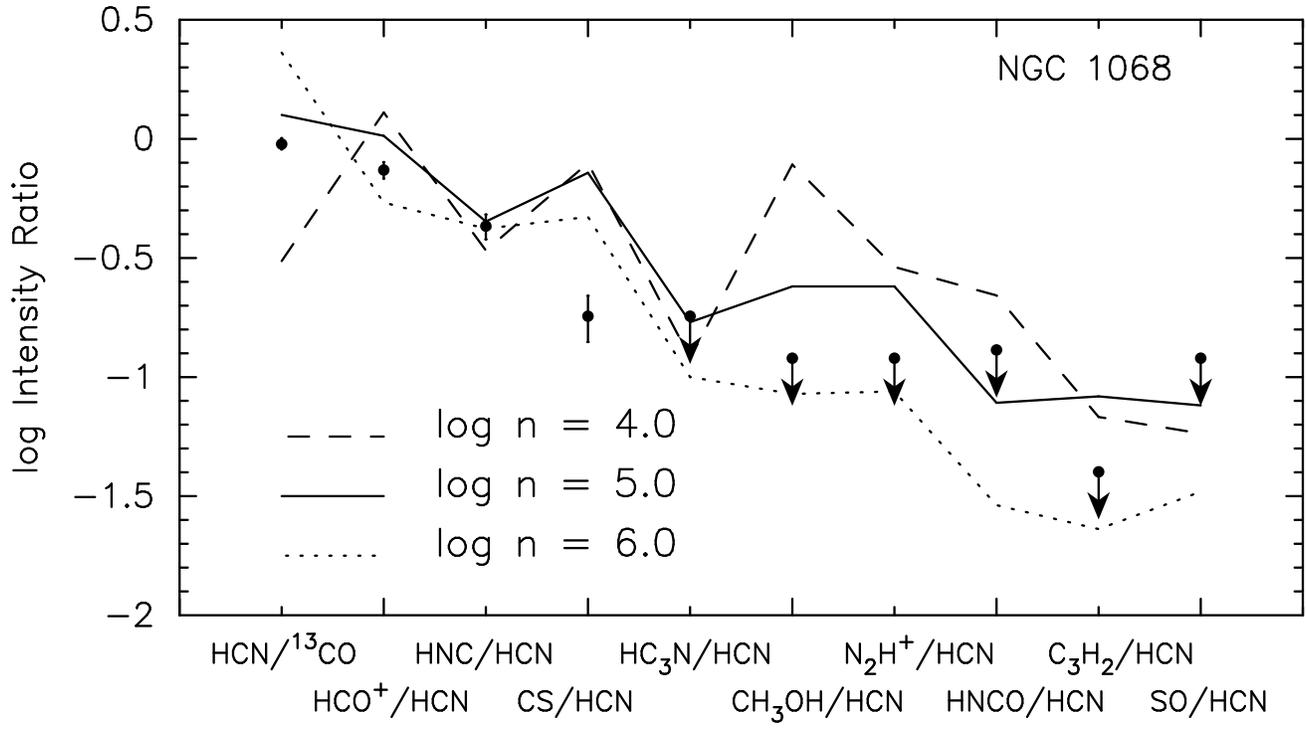}
\figcaption{Same as Figure 3(a) for NGC 1068.}
\end{center}
\end{figure}

\clearpage
\begin{figure}
\begin{center}
\figurenum{4c}
\includegraphics[angle=-90,scale=0.8]{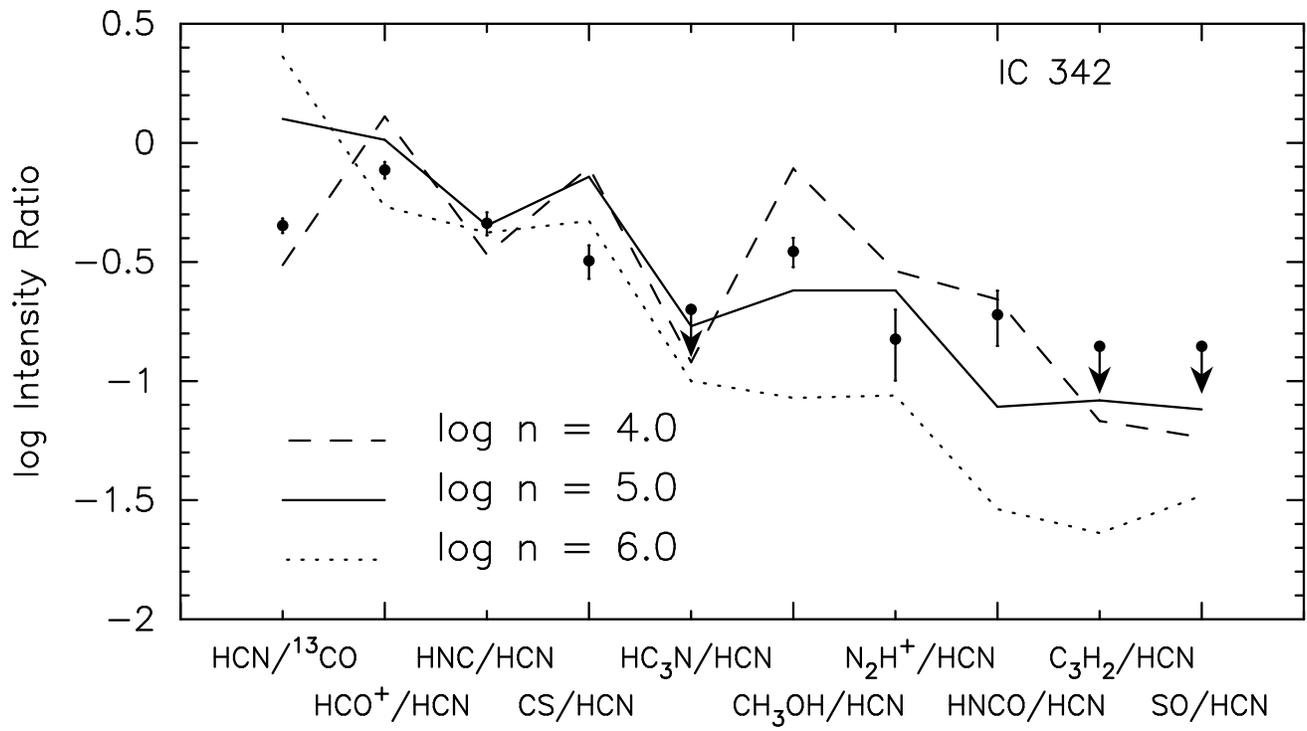}
\figcaption{Same as Figure 3(a) for IC 342.}
\end{center}
\end{figure}

\clearpage
\begin{figure}
\begin{center}
\figurenum{4d}
\includegraphics[angle=-90,scale=0.8]{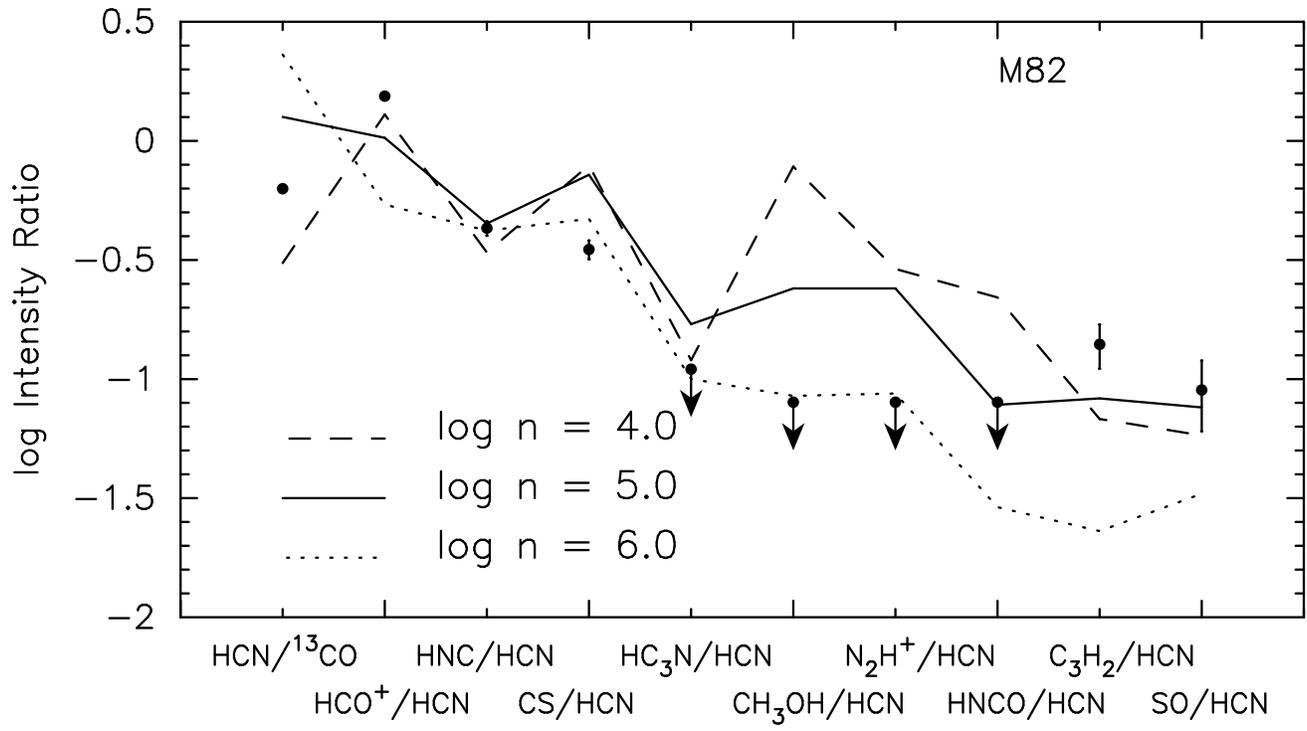}
\figcaption{Same as Figure 3(a) for M82.}
\end{center}
\end{figure}

\clearpage
\begin{figure}
\begin{center}
\figurenum{4e}
\includegraphics[angle=-90,scale=0.8]{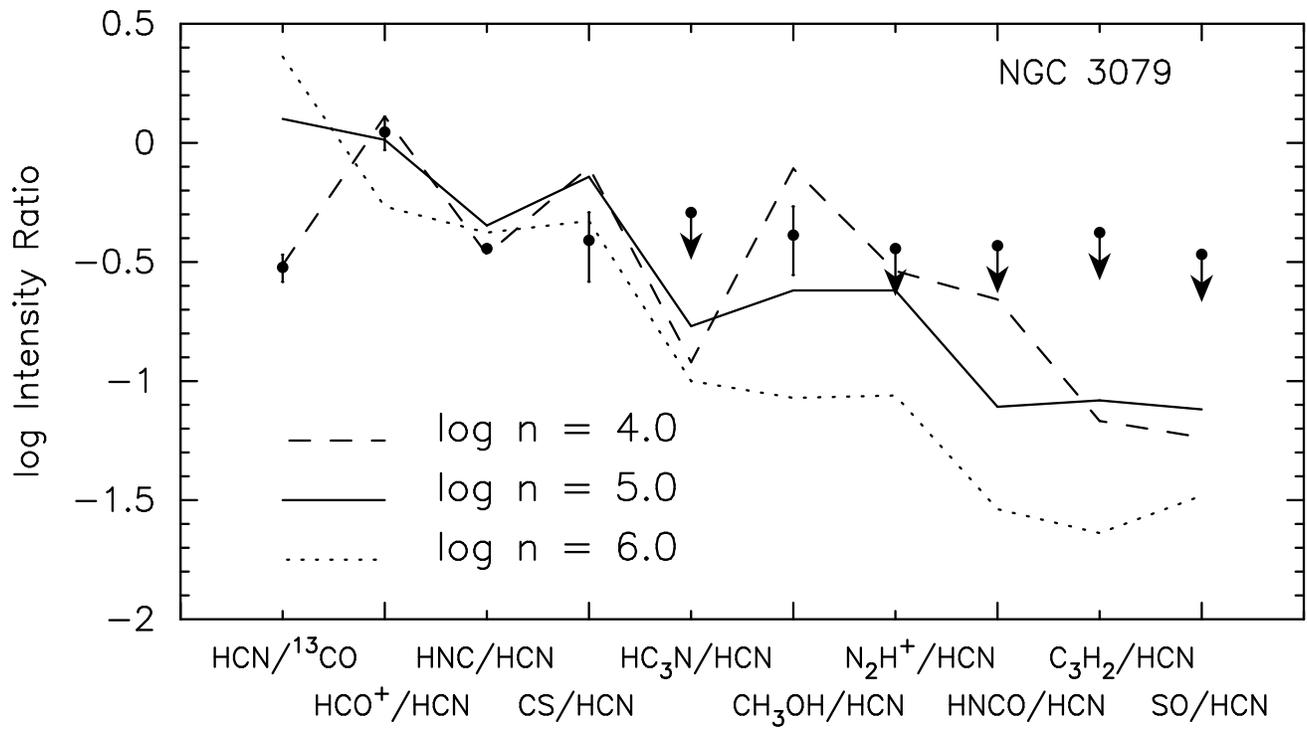}
\figcaption{Same as Figure 3(a) for NGC 3079. }
\end{center}
\end{figure}

\clearpage
\begin{figure}
\begin{center}
\figurenum{4f}
\includegraphics[angle=-90,scale=0.8]{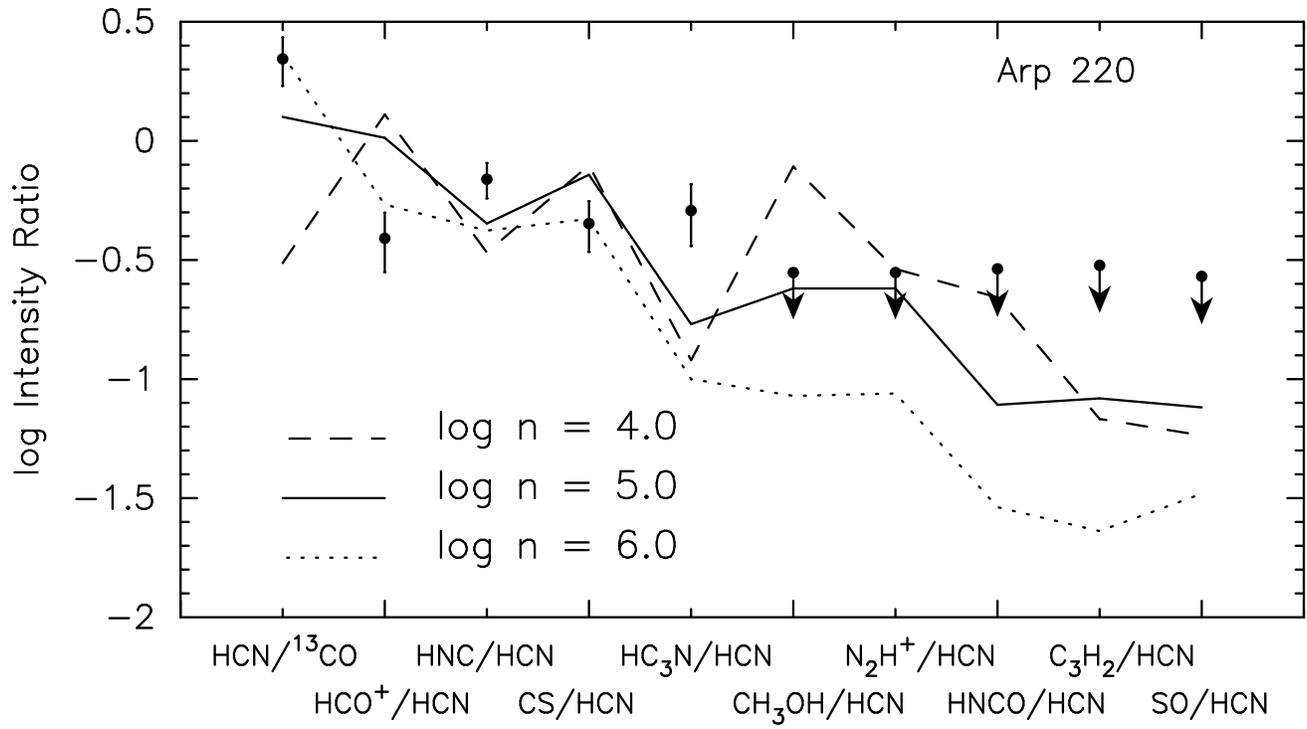}
\figcaption{Same as Figure 3(a) for Arp 220.}
\end{center}
\end{figure}

\clearpage

\begin{deluxetable}{lrrccc}
\tabletypesize{\footnotesize}
\tablewidth{0pt}
\tablecaption{Galaxy Sample and Observing Times}
\tablehead{
\colhead{Galaxy}  & \colhead{$\alpha$(2000)\tablenotemark{1}} & \colhead{$\delta$(2000)\tablenotemark{1}}
& \colhead{Distance} & \colhead{t$_{int}$} & \colhead{Notes} \\
\colhead{}  & \colhead{}  & \colhead{}  & \colhead{Mpc}  & \colhead{hr}}\\
\startdata
NGC 253         & 00 47 33 & -25 17 18 & 3.9   & 3.2  &  Starburst\\
Maffei 2        & 02 41 55 & +59 36 01 & 2.8   & 1.8  & Starburst\\
NGC 1068        & 02 42 40 & -00 00 48 & 14.4  & 3.5 &  Starburst, Seyfert 2 \\
IC 342          & 03 46 48 & +68 05 45 & 3.3   & 8.3  & Starburst \\
M82             & 09 55 52 & +69 40 47 & 3.5   & 4.3  &  Starburst \\
NGC 3079        & 10 01 57 & +55 40 46 & 19.7  & 7.5  & LINER\\
NGC 3690        & 11 28 32 & +58 33 43 & 45    & 9.0  & LIRG, AGN \\
NGC 4258        & 12 18 57 & +47 18 13 & 7.9   & 3.5 & Seyfert 2 \\
Arp 220         & 15 34 57 & +23 30 09 & 76    & 6.0  & ULIRG \\
NGC 6240        & 16 52 58 & +02 24 03 & 101   & 9.7 & LIRG, Seyfert 2 \\
\enddata
\tablenotetext{1}{Units of right ascension are hours, minutes, and seconds, and units of declination are degrees, arcminutes, and arcseconds.}
\end{deluxetable}
\clearpage

\begin{deluxetable}{ccc}
\tabletypesize{\scriptsize}
\tablewidth{0pt}
\tablecaption{Fitted Spectral Lines}
\tablehead{
\colhead{Line Number} & \colhead{Frequency (GHz)} & \colhead{Identification}} \\
\startdata
1  &  74.6446	&   H44 $\alpha$  \\
2  &  79.9127   &   H43 $\alpha$  \\
3  &  81.8815   &   HC$_3$N 9 - 8  \\
4  &  84.5211   &   CH$_3$OH 5(-1) - 4(0) E  \\
5  &  85.3389   &   C$_3$H$_2$ 2(1,2) - 1(0,1)  \\
6  &  85.4557   &   CH$_3$C$_2$H blend  \\
7  &  85.6884   &   H42 $\alpha$  \\
8  &  86.3402   &   H$^{13}$CN  1 - 0  \\
9  &  86.7543   &   H$^{13}$CO$^+$ 1 - 0  \\
10 &  87.0909   &   HN$^{13}$C 1 - 0  \\
11 &  87.3169   &   C$_2$H blend  \\
12 &  87.4020   &   C$_2$H blend  \\
13 &  87.9252   &   HNCO 4(1,3) - 3(1,2)  \\
14 &  88.6318   &   HCN 1-0  \\
15 &  89.1885   &   HCO$^+$ 1 - 0  \\
16 &  90.6635   &   HNC 1-0 \\
17 &  90.9790   &   HC$_3$N 10 - 9  \\
18 &  92.0345   &   H41 $\alpha$  \\
19 &  93.1738   &   N$_2$H$^+$ 1 - 0  \\
20 &  96.4129   &   C$^{34}$S 1 - 0  \\
21 &  96.7414   &   CH$_3$OH blend  \\
22 &  97.9810   &   CS 1 - 0  \\
23 &  99.0230   &   H40 $\alpha$  \\
24 &  99.2999   &   SO 3(2) - 2(1)  \\
25 &  100.0764  &   HC$_3$N 11 - 10 (SO 4(5) - 4(4) blended) \\
26 &  102.5470  &   CH$_3$C$_2$H blend  \\
27 &  106.7374  &   H39 $\alpha$ \\
28 &  108.8939  &   CH$_3$OH 0(0) - 1(-1) E \\
29 &  109.1736  &   HC$_3$N 12 - 11 \\
30 &  109.2522  &   SO 2(3) - 1(2) \\
31 &  109.7822  &   C$^{18}$O 1 - 0 \\
32 &  109.9057  &   HNCO 5(0,5) - 4(0,4)  \\
33 &  110.2014  &   $^{13}$CO 1 - 0  \\
\enddata
\end{deluxetable}

\clearpage

\begin{deluxetable}{ccccccccccc}
\tabletypesize{\scriptsize}
\tablewidth{0pt}
\tablecaption{Fitted Line Intensities and Line Widths}
\tablehead{
\colhead{} & \multicolumn{10}{c}{Line Intensities\tablenotemark{1}} \\
\cline{2-11} \\
\colhead{Line Number} & \colhead{NGC253} & \colhead{Maffei 2} & \colhead{NGC1068} & \colhead{IC342} &
\colhead{M82} & \colhead{NGC3079} & \colhead{NGC3690} & \colhead{NGC4258} & \colhead{Arp220}
& \colhead{NGC6240}}
\startdata
1 & $<$2.2 & $<$2.6 & $<$1.5 & $<$1.9 & $<$1.8 & $<$0.6 & $<$0.9 & $<$1.1 & $<$0.9 & $<$0.6 \\
2 & $<$2.1 & $<$2.5 & $<$1.4 & $<$1.7 & $<$1.7 & $<$0.6 & $<$0.9 & $<$1.0 & $<$0.9 & $<$0.6 \\
3 & 4.7(0.7) & $<$2.5 & $<$1.4 & $<$1.7 & $<$1.7 & $<$0.6 & $<$0.9 & $<$1.0 & $<$0.9 & $<$0.6 \\
4 & $<$2.0 & $<$2.4 & $<$1.4 & $<$1.7 & $<$1.6 & $<$0.6 & $<$0.8 & $<$1.0 & $<$0.9 & $<$0.6 \\
5 & 2.9(0.7) & $<$2.4 & $<$1.4 & $<$1.6 & 2.7(0.5) & $<$0.7 & $<$0.8 & $<$1.0 & $<$0.9 & $<$0.6 \\
6 & $<$2.0 & $<$2.4 & $<$1.4 & $<$1.8 & $<$1.6 & $<$0.7 & $<$0.8 & $<$1.0 & $<$0.9 & $<$0.6 \\
7 & $<$2.0 & $<$2.4 & $<$1.4 & $<$1.6 & $<$1.6 & $<$0.6 & $<$0.8 & $<$1.0 & $<$0.9 & $<$0.5 \\
8 & $<$2.0 & $<$2.4 & $<$1.4 & $<$1.6 & $<$1.6 & $<$0.6& $<$0.8 & $<$1.0 & $<$0.9 & $<$0.5 \\
9 & $<$2.0 & $<$2.4 & $<$1.4 & $<$1.6 & $<$1.6 & $<$0.6 & $<$0.8 & $<$1.0 & $<$0.9 & $<$0.5 \\
10 & $<$2.0 & $<$2.4 & $<$1.4 & $<$1.6 & $<$1.6 & $<$0.6 & $<$0.8 & $<$1.0 & $<$0.9 & $<$0.5 \\
11 & 9.0(0.7) & $<$2.4 & 2.4(0.5) & 2.0(0.5) & 9.5(0.6) & $<$0.8 & $<$0.9 & $<$1.0 & $<$0.9 & $<$0.7 \\
12 & 4.3(0.7) & $<$2.4 & $<$1.4 & $<$1.6 & 4.6(0.6) & $<$0.8 & $<$0.9 & $<$1.0 & $<$0.9 & $<$0.7 \\
13 & 4.6(0.7) & 2.8(0.8) & $<$1.4 & 2.2(0.6) & $<$1.6 & $<$0.6 & $<$0.9 & $<$1.0 & $<$0.8 & $<$0.5 \\
14 & 37.5(0.7) & 8.2(0.8) & 10.8(0.5) & 11.4(0.6) & 20.1(0.6) & 1.6(0.2) & $<$0.8 & $<$1.0 & 2.9(0.3) & $<$0.5 \\
15 & 32.6(0.7) & 7.4(0.8) & 8.0(0.5) & 8.7(0.5) & 31.0(0.6) & 1.8(0.2) & 2.5(0.3) & 1.8(0.3) & 1.1(0.3) & 1.2(0.2) \\
16 & 19.4(0.7) & 5.2(0.8) & 4.7(0.5) & 5.2(0.5) & 8.6(0.5) & $<$0.6 & 1.2(0.3) & $<$1.0 & 2.0(0.3) & $<$0.5 \\ 
17 & 3.3(0.7) & $<$2.3 & $<$1.5 & $<$1.7 & $<$1.6 & $<$0.6 & $<$0.8 & $<$1.0 & $<$0.8 & $<$0.5 \\
18 & $<$2.0 & $<$2.3 & 1.5(0.5) & $<$1.5 & $<$1.6 & $<$0.6 & $<$0.8 & 2.4(0.3) & $<$0.8 & $<$0.5 \\
19 & 5.5(0.6) & $<$2.3 & $<$1.3 & 1.7(0.5) & $<$1.6 & $<$0.6 & $<$0.8 & $<$1.0 & $<$0.8 & $<$0.5 \\
20 & $<$1.9 & $<$2.3 & $<$1.3 & $<$1.6 & $<$1.5 & $<$0.6 & $<$0.8 & $<$0.9 & $<$0.9 & $<$0.5 \\
21 & 7.4(0.6) & 3.0(0.8) & $<$1.3 & 4.0(0.5) & $<$1.5 & 0.6(0.2) & $<$0.8 & $<$0.9 & $<$0.9 & $<$0.5 \\
22 & 14.4(0.6) & 5.4(0.8) & 2.0(0.4) & 3.6(0.5) & 7.1(0.5) & 0.6(0.2) & $<$0.8 & $<$0.9 & 1.3(0.3) & $<$0.5 \\
23 & $<$1.9 & $<$2.2 & $<$1.3 & $<$1.6 & $<$1.5 & $<$0.5 & $<$0.8 & $<$0.9 & $<$0.9 & $<$0.5 \\
24 & $<$1.9 & $<$2.2 & $<$1.3 & $<$1.6 & 1.8(0.5) & $<$0.5 & $<$0.8 & $<$0.9 & $<$0.8 & $<$0.5 \\
25 & 4.4(0.6) & $<$2.2 & $<$1.3 & $<$1.6 & $<$1.5 & $<$0.5 & $<$0.8 & $<$0.9 & 1.1(0.3) & $<$0.5 \\
26 & $<$1.8 & $<$2.2 & $<$1.3 & $<$1.6 & 2.9(0.5) & $<$0.5 & $<$0.8 & $<$0.9 & $<$0.8 & $<$0.5 \\
27 & $<$1.8 & $<$2.2 & $<$1.3 & $<$1.5 & 1.7(0.5) & $<$0.5 & $<$0.7 & $<$0.9 & $<$0.8 & $<$0.5 \\
28 & $<$1.8 & $<$2.1 & $<$1.2 & $<$1.5 & $<$1.4 & $<$0.6 & $<$0.7 & $<$0.9 & $<$0.8 & $<$0.5 \\
29 & $<$1.9 & $<$2.2 & $<$1.4 & $<$1.5 & $<$1.6 & $<$0.9 & $<$0.9 & $<$1.0 & 1.2(0.3) & $<$0.8 \\
30 & $<$1.9 & $<$2.2 & $<$1.4 & $<$1.5 & $<$1.6 & $<$0.9 & $<$0.9 & $<$1.0 & $<$1.0 & $<$0.8 \\
31 & 9.5(0.6) & 2.9(0.7) & 2.2(0.4) & 6.4(0.5) & 7.1(0.5) & 1.6(0.2) & $<$0.8 & $<$0.9 & $<$0.8 & $<$0.6 \\
32 & 2.7(0.6) & $<$2.1 & $<$1.2 & 2.1(0.5) & $<$1.4 & $<$0.7 & $<$0.8 & $<$0.9 & $<$0.8 & $<$0.6 \\
33 & 43.2(0.7) & 22.7(0.8) & 11.4(0.5) & 25.6(0.6) & 31.8(0.5) & 5.3(0.2) &1.2(0.3) & 3.3(0.3) &1.3(0.3) & $<$0.5 \\
\hline
Line Width\tablenotemark{2} & 235(3) & 202(8) & 277(9) & 132(4) & 253(4) & 546(22) & 357(37) & 305(27) & 361(30) & 551(75)
\enddata
\tablenotetext{1}{Intensities given are the peak T$_A^*$ in mK. The format is intensity with the 1 sigma uncertainty in paratheses.  For for non-detections, the 3 sigma upper limit on the line intensity is given.}
\tablenotetext{2}{Line widths are in km s$^{-1}$.  The format is line width with the 1 sigma uncertainty 
in paratheses.}
\end{deluxetable}

\begin{deluxetable}{ccc}
\tabletypesize{\scriptsize}
\tablewidth{0pt}
\tablecaption{Assumed Molecular Abundance Relative to H$_2$}
\tablehead{
\colhead{Molecular Species} & \colhead{Ratio - GMCs\tablenotemark{1}} 
& \colhead{Ratio - NGC 253\tablenotemark{2}}} \\
\startdata
$^{13}$CO  &  2$\times10^{-6}$  & - \\
HCN  &  9$\times10^{-9}$  & 5.0$\times10^{-9}$ \\
HCO$^+$  &  2$\times10^{-9}$  & 1.6$\times10^{-9}$ \\
HNC  &  2$\times10^{-9}$  & 1.0$\times10^{-9}$ \\
CS  &  6$\times10^{-9}$  & 6.2$\times10^{-9}$ \\
N$_2$H$^+$  &  3$\times10^{-10}$ & - \\
HC$_3$N &  4$\times10^{-10}$  & 6.3$\times10^{-9}$ \\
SO  &  1$\times10^{-9}$  & 1.3$\times10^{-9}$ \\
CH$_3$OH  &  8$\times10^{-9}$ & 1.3$\times10^{-9}$ \\
C$_3$H$_2$ & 5$\times10^{-10}$ & 5.0$\times10^{-10}$ \\
HNCO  &  1$\times10^{-9}$  & 1.6$\times10^{-9}$ \\
\enddata
\tablenotetext{1}{Abundances from \citet{ber97}}
\tablenotetext{1}{Abundances from \citet{mar06}}
\end{deluxetable}

\begin{deluxetable}{lcccccc}
\tabletypesize{\scriptsize}
\tablewidth{0pt}
\tablecaption{Mass and Beam Filling Fraction}
\tablehead{
\colhead{Galaxy} & \colhead{D$_B$} & \colhead{f$_{area}$} & \colhead{Mass from \thco} & 
\colhead{S$_{CO}$\tablenotemark{1}} & \colhead{Mass from CO} & \colhead{Dense Gas Fraction} \\
\colhead{} & \colhead{kpc} & \colhead{} & \colhead{M$_\sun$} & \colhead{Jy km s$^{-1}$}
& \colhead{M$_\sun$}} \\
\startdata
NGC 253 & 0.9 & 0.26 & 2.7$\times10^8$ & 1.1$\times10^4$  & 1.2$\times10^9$ & 0.23\\
Maffei 2 & 0.6 & 0.11 & 6.2$\times10^7$ & - & - & - \\
NGC 1068 & 3.3  & 0.08  & 1.1$\times10^9$ & 2.3$\times10^3$  & 1.8$\times10^9$ & 0.61 \\
IC 342  & 0.8  & 0.08  & 6.2$\times10^7$ & 1.5$\times10^3$  & 1.2$\times10^8$ & 0.52 \\
M82   &  0.8   &  0.19  & 1.7$\times10^8$ & 6.7$\times10^3$  & 5.8$\times10^8$ & 0.29 \\
NGC 3079  &  4.5  & 0.07  & 2.0$\times10^9$ & 1.6$\times10^3$  & 4.4$\times10^9$ & 0.45 \\
NGC 3690  &  10.4   & 0.01  &  1.5$\times10^9$ & 4.5$\times10^2$  & 6.5$\times10^9$ & 0.23 \\
NGC 4258  &  1.8  &  0.02  & 1.1$\times10^8$ & 3.0$\times10^2$  & 1.3$\times10^8$ & 0.85 \\
Arp 220  &  17.5  & 0.01  &  4.7$\times10^9$ & 4.0$\times10^2$ & 1.6$\times10^{10}$ & 0.29 \\
NGC 6240 &  23.2 & $<$0.01  & $<$4.9$\times10^9$  & 2.4$\times10^2$ & 1.7$\times10^{10}$ & $<$0.29 \\
\enddata
\tablenotetext{1}{From \citet{you95}}
\end{deluxetable}


\begin{thebibliography}{}
\bibitem[Aalto et al.(1997)]{aal97} Aalto, S., Radford, S. J. E., Scoville, N.Z. \&
Sargent, A. I. 1997, \apjl, 475, L107
\bibitem[Aalto et al.(2002)]{aal02} Aalto, S., Polatidis, A.G., H\"uttemeister, S. \&
Curran, S.J., 2002, \aap, 381, 783
\bibitem[Aalto et al.(2007)]{aal07} Aalto, S., Monje, R. \& Mart\'{i}n, S. 2007, \aap, 475, 479
\bibitem[Aalto(2008)] {aal08} Aalto, S. 2008, \apss, 313, 273
\bibitem[Aladro et al.(2011)]{ala10} Aladro, R., Mart\'{i}n-Pintado, J., Mart\'{i}n, S., 
Mauersberger, R. \& Bayet, E. 2011, \aap, 525, 89
\bibitem[Baan et al.(2008)] {baa08} Baan, W. A., Henkel, C., Loenen, A. F., Baudry, A.
\& Wiklind, T., 2008, \aap, 477, 747
\bibitem[Bally et al.(1987)] {bal87} Bally, J., Stark, A., Wilson, W. \& Henkel, C. 1987, 
\apjs, 65, 13
\bibitem[Bergin et al.(1997)]{ber97} Bergin, E. A., Ungerechts, H., Goldsmith, P.F., 
Snell, R.L., Irvine, W.M. \& Schloerb, F.P., 1997, \apj, 482, 267
\bibitem[Bayet et al.(2009)]{bay09} Bayet, E., Aladro, R., Mart\'{i}n, S., Viti, S.,
\& Mart\'{i}n-Pintado, J. 2009, \apj, 707, 126
\bibitem[Carpenter et al.(1995)] {car95} Carpenter, J.M., Snell, R.L. \& Schloerb, F.P. 1995,
\apj, 445, 246
\bibitem[Churchwell et al.(1984)]{chu84} Churchwell, E., Nash, A.G. \& Walmsley, C. M.,
1984, \apj, 287, 681
\bibitem[Cummins et al.(1986)] {cum86} Cummins, S. E., Linke, R. A., \& Thaddeus, P. 1986,
\apjs, 60, 819
\bibitem[Erickson et al.(2007)]{eri07} Erickson, N., Narayanan, G., Goeller, R., \&
Grosslein, R. 2007, ASP Conf. Ser. 375, {\it From Z-Machines to ALMA: (Sub)millimeter Spectroscopy
of Galaxies}, ed. A. J. Baker, J. Glenn, A. I. Harris, J.G. Mangum, M.S. Yun (San Francisco, CA: ASP), 71
\bibitem[Fuente et al(2005)]{fue05} Fuente, A., Garc\'{i}a-Burillo, S., Gerin,M., Teyssier,
D., Usero, A., Rizzo, J.R. \& De Vicente, P. 2005 \apjl, 619, L155
\bibitem[Gao \& Solomon(2004)]{gao04} Gao, Y. \& Solomon, P. M., 2004, \apj, 606, 271
\bibitem[Goldsmith et al.(1981)]{gol81} Goldsmith, P.F., Langer, W.D., Ellder, J., 
Irvine, W. \& Kollberg, E., 1981, \apj, 249, 524 
\bibitem[Graci\'{a}-Carpio et al.(2006)]{gra06} Graci\'{a}-Carpio, J., Garc\'{i}a-Burillo,
S., Planesas, P., \& Colina, L., 2006 \apjl, 640, L135
\bibitem[Garc\'{i}a-Mar\'{i}n et al.(2006)] {gar06} Garc\'{i}a-Mar\'{i}n, M., Colina, L., 
Arribas, S., Alonso-Herrero, A. \& Mediavilla, E. 2006, \apj, 650, 850
\bibitem[Greve et al.(2009)]{gre09} Greve, T. R., Papadopoulos, P. P., Gao, R. \& Radford, 
S. J. E. 2009, \apj, 692, 1432
\bibitem[H\"{u}ttemeister et al.(1997)] {hut97} H\"{u}ttemeister, S., Mauersberger, R.
\& Henkel, C. 1997, \aap, 326, 59
\bibitem[Imanishi et al.(2007)] {ima07} Imanishi, M., Nakanishi, K., Tamura, Y., Oi, Nagisa, \&
Kohno, Kotaro, 2007, \aj, 134, 2366
\bibitem[Iono et al.(2007)]{ion07} Iono, D., Wilson, C.D., Takakuwa, S., Yun, M. S., Petitpas, G.R.,
Peck, A. B., Ho, P.T.P., Matsushita, S., Pihlstrom, Y. M. \& Wang, Z. 2007, \apj, 659, 283
\bibitem[Johansson et al.(1984)]{joh84} Johansson, L. E. B., Andersson, C., Ellder, J., Friberg, P.,
Hjalmarson, A., Hoglund, B., Irvine, W. M., Olofsson, H. \& Rydbeck, G. 1984, \aap, 130, 227
\bibitem[Johansson et al.(1985)]{joh85} Johansson, L. E. B., Andersson, C., Elder, J., Friberg, P.,
Hjalmarson, A., Hoglund, B., Olofsson, H., Rydbeck, G. \& Irvine, W. M. 1985, \aaps, 60, 135
\bibitem[Juneau et al.(2009)]{jun09} Juneau, S., Narayanan, D. T., Moustakas, J., Shirley,
Y. L., Bussman, R. S., Kennicutt, R. C., Vanden Bout, P. A. 2009, \apj, 707, 1217
\bibitem[Karachentsev(2005)]{kar05} Karachentsev, I, D. 2005, \apj, 129, 178
\bibitem[Kenney \& Young(1989)]{ken89} Kenney, J.D.P., Young, J.S. 1989, \apj, 344, 171
\bibitem[Kohno et al.(2001)] {koh01} Kohno, K., Matsushita, S., Vila-Vilar\'{o}, B., 
Okumura, S. K., Shibatsuka, T., Okiura, M., Ishizuki, S., Kawabe, R., 2001, ASP Conf. Ser. 249,
{\it The Central Kiloparsec of Starbursts and AGN: The La Palma Connection},
ed. J. H. Knapen, J. E. Beckman, I. Shlosman, \& T. J. Mahoney (San Francisco, CA: ASP), 672
\bibitem[Knudsen et al.(2007)] {knu07} Knudsen, K. K., Walter, F., Weiss, A., Bollatto, A.,
Riechers, D. A. \& Menten, K. 2007, \apj, 666, 156
\bibitem[Krause et al.(2007)]{kra07} Krause, M, Fendt, C. \& Neininger, N. 2007, \aap, 467, 1037
\bibitem[Lee et al.(1990)]{lee90} Lee, Y., Snell, R. \& Dickman, R. 1990, \apj, 355, 536
\bibitem[Lepp \& Dalgarno(1996)]{lep96} Lepp, S, \& Dalgarno, A., 1996, \aap, 306, L21
\bibitem[Loenen(2009)]{loe09} Loenen, E., 2009, Ph.D. thesis, Univ. of Groningen
\bibitem[Maloney et al.(1996)]{mal96} Maloney, R.R., Hollenbach, D.J. \& Tielens, A.G.G.M. 
1996, \apj, 466, 561
\bibitem[Mart\'{i}n et al.(2005)]{mar05} Mart\'{i}n, S., Mart\'{i}n-Pintado, J., Mauersberger,
R., Henkel, C. \& Garc\'{i}a-Burillo, S. 2005, \apj, 620, 210
\bibitem[Mart\'{i}n et al.(2006)]{mar06} Mart\'{i}n, S., Mauersberger, R., Mart\'{i}n-Pintado,
J., Henkel, C., \& Garc\'{i}a-Burillo, S. 2006, \apjs, 164, 450
\bibitem[Mart\'{i}n et al.(2008)]{mar08} Mart\'{i}n, S., Requena-Torres, M.A., Mart\'{i}n-Pintado,
J. \& Mauersberger, R. 2008, \apj, 678, 245
\bibitem[Meier \& Turner(2005)]{meie05} Meier, D. S. \& Turner, J. L. 2005, \apj, 618, 259
\bibitem[Meijerink \& Spaans(2005)]{mei05} Meijerink, R. \& Spaans, M., 2005, \aap, 436, 397
\bibitem[Meijerink et al.(2007)]{mei07} Meijerink, R., Spaans, M. \& Israel, F., 2007, \aap, 461, 793
\bibitem[Naylor et al.(2010)]{nay10} Naylor, B.J., Bradford, C.M., Aquirre, J.E., Bock, J.J.,
Earle, L., Glenn, J., Inami, H., Kamenetzky, J., Maloney, P.R., Matsuhara, H., Nguyen, H.T., \&
Zmuidzinas, J. 2010, \apj, 722, 558
\bibitem[Omont(2007)] {omo07} Omant, A., 2007, Rep. Prog. Phys., 70, 1099
\bibitem[Paglione et al.(1998)]{pag98} Paglione, T., Jackson, J., Bolatto, A., \& Heyer, M. 1998,
\apj, 493, 680
\bibitem[Paglione et al.(2001)]{pag01} Paglione, T., Wall, W., Young, J., Heyer, M.,
Richard, M., Goldstein, M., Kaufman, Z., Nantais, J., \& Perry. G. 2001, \apjs, 135, 183.
\bibitem[Penzias et al.(1972)] {pen72} Penzias, A. A., Jefferts, K. B., Wilson, R. W.,
Liszt, H. S., \& Solomon, P. M., 1972, \apjl, 178, L35
\bibitem[Schilke et al.(1992)]{sch92} Schilke, P., Walmsley, C.M., Pineau Des Forets. G.,
Roueff, E., Flower, D.R. \& Guilloteau, S., 1992, \aap, 256, 595
\bibitem[Schinnerer et al.(2000)]{sch00} Schinnerer, E., Eckart, A., Tacconi, L.J., Genzel,
R. \& Downes, D., 2000, \apj, 533 850
\bibitem[Schloerb(2008)]{sch08} Schloerb, F. P. 2008, Proc. SPIE, 7012
\bibitem[Sch\"{o}ier et al.(2005)]{sch05} Sch\"{o}ier, F.L., Van der Tak, F.F.S., 
van Dishoeck, E.F. \& Blaqck, J. H., 2005, \aap, 432, 369
\bibitem[Snell et al.(1984)] {sne84} Snell, R. L., Goldsmith, P. F., Erickson, N. R., Mundy, L. G.
\& Evans, N. J. II 1984, \apj, 276 625 
\bibitem[Tideswell et al.(2010)]{tid09} Tideswell, D.M., Fuller, G.A., Millar, T.J. \&
Markwick, A.J., 2010, \aap, 510, 85
\bibitem[Turner(1989)]{tur89} Turner, B. 1989, \apjs, 70, 539
\bibitem[Van der Tak et al.(2007)]{van07} Van der Tak, F.F.S., Black, J.H., Sch\"{o}ier, F.L.,
Jansen, D.J. \& van Dishoeck, E.F., 2007,\aap, 468, 627
\bibitem[Wilson(1999)]{wil99} Wilson, T.L., 1999, Rep. Prog. Phys., 62, 143
\bibitem[Wilson \& Rood(1994)]{wil94} Wilson, T.L. \& Rood, R. 1994, \araa, 32, 191
\bibitem[Young \& Scoville(1991)] {you91} Young, J.S. \& Scoville, N.Z. 1991, \araa, 29, 581
\bibitem[Young et al.(1995)]{you95} Young, J.S. et al. 1995, \apjs, 98, 219


\end{thebibliography}
\end{document}